\documentclass[useAMS,usegraphicx,usenatbib,referee]{biom}

\usepackage{subcaption}
\usepackage{dsfont}
\usepackage{amssymb}
\usepackage{amsmath}
\usepackage{hyperref}
\usepackage[xcdraw,table]{xcolor}
\usepackage{multirow}
\usepackage{algorithm}
\usepackage{algorithmicx}
\usepackage{algpseudocode}
\newcommand{\qu}[1]{``#1''}
\newcommand{\bv}[1]{\bmath{#1}}
\newcommand{\sigsq}{\sigma^2}
\newcommand{\tausq}{\tau^2}
\newcommand{\x}{\bv{x}}

\newcommand{\X}{\bv{X}}
\newcommand{\w}{\bv{w}}

\newcommand{\Ybar}{\bar{Y}}
\newcommand{\YbarT}{\Ybar_T}
\newcommand{\YbarC}{\Ybar_C}

\newcommand{\betaT}{\beta_T}

\newcommand{\nr}{n_{R}}
\newcommand{\nrc}{n_{R,C}}
\newcommand{\nrt}{n_{R,T}}

\newcommand{\Dbar}{\bar{D}}
\newcommand{\SsqR}{S^2_R}
\newcommand{\YbarRT}{\Ybar_{R,T}}
\newcommand{\YbarRC}{\Ybar_{R,C}}

\newcommand{\SsqDbar}{S^2_{\Dbar}}

\newcommand{\YbarRTMinusYbarRC}{\YbarRT - \YbarRC}

\newcommand{\sigsqR}{\sigsq_R}
\newcommand{\sigsqDbar}{\sigsq_{\bar{D}}}

\DeclareMathOperator*{\argmin}{arg\,min~}
\newcommand{\Sinv}{\bv{S}^{-1}}
\newcommand{\parens}[1]{\left(#1\right)}
\newcommand{\inverse}[1]{\parens{#1}^{-1}}
\newcommand{\braces}[1]{\left\{#1\right\}}
\newcommand{\bracks}[1]{\left[#1\right]}
\newcommand{\ceiling}[1]{\lceil#1\rceil}
\newcommand{\angbrace}[1]{\left<#1\right>}
\newcommand{\squared}[1]{\parens{#1}^2}
\newcommand{\overtwo}[1]{\frac{#1}{2}}

\newcommand{\convp}{~{\buildrel p \over \rightarrow}~}
\newcommand{\convd}{~{\buildrel \mathcal{D} \over \longrightarrow}~}

\newcommand{\oneover}[1]{\frac{1}{#1}}

\newcommand{\epsilonrv}{\mathcal{E}}
\newcommand{\normnot}[2]{\mathcal{N}\parens{#1,\,#2}}

\newcommand{\stdnormnot}{\normnot{0}{1}}
\newcommand{\bernoulli}[1]{\rmn{Bern}\parens{#1}}
\newcommand{\iid}{~{\buildrel iid \over \sim}~}
\newcommand{\sigsqe}{\sigsq_e}

\newcommand{\beqn}{\begin{eqnarray*}}
\newcommand{\eeqn}{\end{eqnarray*}}
\newcommand{\bneqn}{\begin{eqnarray}}
\newcommand{\eneqn}{\end{eqnarray}}

\title[Better Matching in Sequential Experiments]{A Matching Procedure for Sequential Experiments that Iteratively Learns which Covariates Improve Power}

\author
{Adam Kapelner$^{*}$\email{kapelner@qc.cuny.edu} \\
Department of Mathematics, Queens College, City University of New York, \\ 65-30 Kissena Blvd, Queens, NY, 11367 
\and 
Abba Krieger$^{**}$\email{krieger@wharton.upenn.edu} \\
Department of Statistics, The Wharton School at the University of Pennsylvania, \\ 3730 Walnut St, Philadelphia, PA, 19104}

\pubyear{Year}
\artmonth{Month}

\begin{document}

\begin{abstract}
We propose a dynamic allocation procedure that increases power and efficiency when measuring an average treatment effect in sequential randomized trials exploiting some subjects' previous assessed responses. Subjects arrive sequentially and are either randomized \textit{or} paired to a previously randomized subject and administered the alternate treatment. The pairing is made via a dynamic matching criterion that iteratively learns which specific covariates are important to the response. We develop estimators for the average treatment effect as well as an exact test. We illustrate our method's increase in efficiency and power over other allocation procedures in both simulated scenarios and a clinical trial dataset. An \texttt{R} package \qu{\texttt{SeqExpMatch}} for use by practitioners is available.
\end{abstract}


\maketitle

\section{Introduction}\label{sec:introduction}

We consider the classic two-arm sequential experiment where $t = 1 \ldots n$ subjects enter sequentially and must be assigned immediately to one of two experimental groups, a \emph{treatment} group (denoted T) and a \emph{control} group (denoted C). The goal is to estimate and test a causal treatment effect. This experiment has a clearly defined outcome of interest, $y$ (also called the response or endpoint), collected sequentially and we scope our discussion to $y$ being continuous and uncensored (incidence and survival endpoints are left to future work). The sequential setting (or \qu{online} setting) is the majority of clinical trials \citep[Myth \#1]{Senn2013}, a \$40 billion industry annually, popular in crowdsourced-Internet experimentation \citep{Horton2011, Chandler2013} and also popularly run by Facebook, Amazon and other large technology companies. And in many of these settings the responses are collected rapidly and available in sequence.

\qu{Gold standard measurements} of causal effects are done via sequentially assigning subjects to treatment groups randomly. The justification for randomization is elegantly summed up by \citet[page 245]{Cornfield1959} who wrote that known and unknown covariate differences among the T and C groups will be small and the bounds on these differences are known. 

%there are two downsides that negatively impact estimation efficiency and experimental power. 
What does \qu{randomly} mean precisely in a sequential experiment? The naive sequential assignment is via independent fair coin flips termed the \emph{Bernoulli Trial} \citep[Chapter 4.2]{Imbens2015}. This naive design is unbiased but has downsides that negatively impact estimation efficiency and experimental power. First, the number of assignments to the treatment group is random and will likely be different from the number of assignments to the control group. Second, the covariate differences between treatment groups are anything but small especially for modest sample sizes. Consider each subject to have $p$ covariates measured at the time they entire the trial (and hence are fixed). If they are drawn from independent processes, then the probability of at least one covariate exhibiting a statistically significant mean difference between the T and C groups is $1 - (1 - \alpha)^p$ where $\alpha$ is the level of significance. For a typical clinical trial with 15 covariates and $\alpha = 5\%$, there is more than a 50\% chance that at least one covariate has a statistically significant mean difference, a phenomenon that does not disappear with larger sample sizes.

Solving the first problem of unequal number of treatments and controls is trivial by employing one of the $\binom{n}{n/2}$ allocations vectors that have an equal number of treatments and controls. Designs that do so are termed \emph{forced balance} procedures \citep[Chapter 3.3.]{Rosenberger2016}, the simplest of which employs one of these allocation vectors randomly, a design termed either \emph{complete randomization} or the term we prefer herein, \emph{balanced completely randomized design} (BCRD) as in \citet{Wu1981}. BCRD is considered a \emph{restricted} design since it does not allow for all possible allocations as in the Bernoulli trial, but this restriction is very minor.

One can argue that the second consideration is not important to address as after employing BCRD, the data can be analyzed with a linear regression estimator that can adjust for even severe covariate imbalances. This is true if the model is truly linear; the regression estimator is very difficult to improve upon using experimental design \citep[Section 2.3.2]{Kapelner2020}. However, real-world models exhibit non-linearities and interactions. Designs tailored for the real-world setting can yield significant power improvement and employing the regression estimator in conjunction amplifies this power improvement (we observe this in Section~\ref{sec:simulations}).

\subsection{Non-Sequential Designs}

To discuss solutions to covariate imbalances, we first discuss how this problem is addressed in the non-sequential setting where all covariate values of the $n$ subjects are known prior to treatment allocations (the setting where the field of experimental design historically evolved). Covariate imbalance was immediately noted by Fisher who (ironically) recommends against complete randomization: \citet[page 251]{Fisher1925} wrote that \qu{it is still possible to eliminate much of the \ldots heterogeneity, and so increase the accuracy of our [estimator], by laying restrictions on the order in which the strips are arranged}. Here, he introduces \emph{blocking}, an experimental design strategy that is widely in use today. Blocking is a more \emph{restricted} design than BCRD as it allows for even fewer possible allocations. 

Another strategy discards assignments if they do not meet a threshold of covariate imbalance. Upon discarding, a new assignment is drawn from BCRD. This restricted design is called rerandomization, a heuristic agreed upon by both Student and Fisher but only investigated recently \citep{Morgan2012}.

The extreme restricted designs of direct minimization of a covariate balance function using mixed integer optimization methods has also recently been explored \citep{Bertsimas2015, Kallus2018}. However, severe restrictions that are more deterministic than random violates Cornfield's reason for randomization above (the unknown covariate differences should be small) This violation can outweigh the benefit of reducing the known covariate differences resulting in lower estimation efficiency \citep{Kapelner2020} and lower experimental power \citep{Krieger2020}.

Another method is pairwise matching \citep{Greevy2004}, a classical design that dates back to \citet{Student1931} who recommended performing the $n = 20,000$ children Lanarkshire Milk Experiment on exclusively identical twin pairs. \citet{Kallus2018} proves that for response models with the Lipschitz norm with distance function $d$, maximal efficiency is obtained for the difference-in-means estimator when $d$ of the bipartite matches is minimized. This means that allocating via pairwise matching is a theoretically-grounded robust procedure for a flexible class of response functions.

\subsection{Sequential Designs}

In sequential experiments, the designs are restricted \emph{adaptively}. This means that the function that assigns subject $t$ to the T or C group, is dynamic. This \emph{dynamic allocation} can make use of previous assignments, covariates, and/or responses \citep{HuRosenberger2006} and can adapt to any changes to the experiment while it is underway \citep{Chow2008}. These adaptations can be highly elaborate, all in an effort to decrease estimator variance and increase experimental power. 

Typical sequential designs adapt the non-sequential designs to the sequential setting. The problem of unequal arms can be naively solved by forcing balance with BCRD. This is not typically used in practice as there is risk of a nefarious experimenter who, knowing the order of allocations, may choose his subjects to enter in an order that biases the results to his favor. \citet{Efron1971} provides a medium design balancing the allocation uncertainty of the Bernoulli trial and the forced balance of BCRD with subject assignments via a biased coin (probability not 50\%). An adaptation of Fisher's blocking to the sequential setting is the widely used \emph{minimization} procedure developed by \citet{Pocock1975}. This sequential blocking design minimizes imbalances within the blocks upon each new subject's entrance. Minimization requires arbitrary functions to tailor the imbalances and then employs a biased coin to randomize assignments. Choices of these arbitrary functions are still being investigated \citep{Han2009}. Sequential blocking shares the downsides of non-sequential blocking: the number of blocks grows exponentially with $p$ and the bins of continuous covariates are arbitrarily constructed. The adaptation of non-sequential rerandomization has been explored recently in \citet{Zhou2018}. 

A different approach assumes a linear additive treatment response model allowing for a computation of the proportional variance of the treatment effect estimator. This computation is done at each $t$ for either putative assignment T and C. Normalizing these two inverse proportional variances by their sum yields probabilities of assignment to T and C for subject $t$. This is the biased coin algorithm of \citet[Equation 4]{Atkinson1982} which combines the balancing concept of \citet{Pocock1975} and the biased coin concept of \citet{Efron1971}. \citet{Wiens2005} extends this approach by considering more general response models that include heteroskedasticity, interactions and misspecification and flips the coin with probability inversely proportional to the resulting mean squared error to provide more robust probabilities for the assignments.

Adapting numerical optimization to the sequential setting has a prolific recent literature. Examples can be found in \citet{Bertsimas2019, Bhat2020} but their designs are deterministic which (1) exposes a risk of unbalanced unmeasured covariates and (2) makes the use of a randomization test impossible. \citet{Bertsimas2019} optimize on-the-fly and their allocations are random as they impute values of future subjects' covariates. Finally, pairwise matching was adapted to the sequential setting in \citet{Kapelner2014} where subjects are both randomized and paired on-the-fly. The work herein extends this design and thus henceforth we refer to it as \emph{KK14}.

All these above methods are termed \emph{covariate-adaptive randomization} \citep[Chapter 9]{Rosenberger2016} as the dynamic adaptation is based on previous subjects' covariate measures. In many sequential clinical trials, it is not uncommon for some subjects to have their endpoint $y$ assessed before others enter and are even allocated to a treatment. These previous subjects' responses can be used to influence future allocation. For example, if $y$ is found to be heteroskedastic, it is more efficient to assign more subjects to the group with higher outcome variance. Using $y$ during the experiment is called response adaptive randomization \citep[Chapter 10]{Rosenberger2016}.

Covariate-adjusted response-adaptive (CARA) designs use both covariates and responses \citep[Chapter 1.7]{Rosenberger2016}. Notable methods are \citet{Atkinson2005} who directly optimize the variance of the treatment effect in the context of a linear regression model based on the concept of $D_A$ optimality (for asymptotic properties, see \citealt{Zhang2007}) and \citet{Sverdlov2013} who uses sequential maximum likelihood to create a different biased coin randomization for each subject. 

\subsection{Our Design}

These two abovementioned CARA methods assumed somewhat strict parametric models for the response (such as linearity). Here, we assume a general method and develop a robust design based on matching.%Also, one can assign subjects differently based on circumstances such as failure or noncompliance \citep{Lei2012}. 

%These previous methods either assume a linear model or seek greater observed covariate mean balance because this implies greater efficiency of the treatment effect estimator. However, this is true only if the response is linear and homoskedastic \citep{Rosenberger2008}. In fact, the optimal balance function is tied to which class the response model belongs, or more precisely, the norm put on the response function. For the worst case scenario of the response function, \citet{Kallus2018} proves (for non-sequential trials) that for linear response models, the balance function that should be minimized for maximum treatment estimator efficiency is the Mahalanobis distance, for the Lipschitz norm, the closeness of bipartite matches, for the sup norm, blocking and more. If no assumption is made on the response function, a \qu{no free lunch} result emerges: the minimax optimal design is complete randomization.

%No such theorems are known in the sequential setting. 

In the sequential matching design of KK14, matches were created using covariate distances ignoring the individual covariates' relationships with the response. This creates improved covariate imbalance, but matches that are unfocused on improving estimation efficiency and experimental power, the main goal of restricted experimental design. 

In this paper, we introduce a CARA design which matches subjects on-the-fly using previous subjects' covariates and responses. This information allows us to iteratively detect differential importance of each covariate within the response function. Although the investigator may know a priori how to pare down to a subset of covariates that are most likely to matter for the response, they likely will not know the differential importances among these covariates; and these are the importances we estimate. Additionally, since some of the covariates can be mutually collinear, we also propose a procedure to mitigate double-counting in the importances among the $p$ covariates. We then update the distance function $d$ using these estimated covariate importance weights. 

This new covariate-weighted distance function dramatically improves the quality of sequentially paired matches resulting in more accurate treatment effect estimates and higher experimental power. Our design neither suffers from the risk that a researcher can guess future assignments nor do our assignments exhibit any time trend even if the order of recruitment is related to subject severity. Section \ref{sec:methods} covers our model assumptions, previous work and our new methodology in detail. Section \ref{sec:results} demonstrates our procedure's dramatic improvements over using both clinical data and simulated data and Section~\ref{sec:discussion} concludes and offers future directions.

\section{The Algorithm, Estimation, and Testing}\label{sec:methods}

\subsection{Model Assumptions}\label{subsec:problem_formulation}

There are $n$ total subjects and $t = 1, \ldots, n$ indexes the order in which the subject enters the experiment. Each subject that enters has $p$ covariate measurements (recorded immediately) denoted by $\x_t := \bracks{x_{t1}, \ldots, x_{tp}}$ which are either continuous or binary. In the sequential setting we consider, the subjects must be assigned to one of the two experimental groups immediately upon entry. We let the vector $\w := \bracks{w_1, \ldots, w_n}^\top \in \braces{0,1}^n$ denote all $n$ subjects' assignments where 1 indicates T and 0 indicates C. We assume all previous subjects' covariates, $\x_1, \ldots, \x_{t-1}$, and their corresponding quantitative responses, $y_1, \ldots y_{t-1}$ are known, and this historical information is used to allocate $w_t$.

We assume the same response model as in Equation 1 of KK14, i.e. independent observations, an additive treatment effect, a possibly non-linear covariates effect, normal and homoskedastic errors and fixed covariates:

\bneqn\label{eq:response_model}
Y_t = \beta_T w_t + z_t + \epsilonrv_t, ~~  z_t := f(\x_t) ~~\text{and}~~ \epsilonrv_t \iid \normnot{0}{\sigsqe}.
\eneqn

\noindent Equation~\ref{eq:response_model} above assumes a \emph{population sampling model} where randomness in the responses comes from both the allocation $w_t$ and the noise term $\epsilonrv_t$. This is a strong assumption and is sometimes unjustified in a clinical trial context \citep[Chapter 6.2]{Rosenberger2016}. A weaker assumption is the \emph{Neyman randomization model} \citep[Chapter 6.3]{Rosenberger2016} where the randomness only comes from the allocation,

\bneqn\label{eq:response_model2}
Y_t = \beta_T w_t + z_t.
\eneqn

\noindent This distinction will become important when we construct our estimator $B_T$ for $\beta_T$ and use it to construct frequentist confidence intervals and run hypothesis tests.

\subsection{The Algorithm}\label{subsec:algorithm}

The first $t_0 \geq p$ subjects (where $t_0$ is a tuning hyperparameter) are randomized as in the Bernoulli Trial and the collecion of these $t_0$ subjects initializes the set we call the \qu{reservoir} (see lines 2-3 of Algorithm~\ref{alg:matching}). For each subject that enters later than $t_0$ we attempt to match to a subject in the reservoir (lines 4-15). If the reservoir is empty, they are randomized as in the Bernoulli trial (line 2). 

Our design exploits the previous responses $y_1, \ldots y_{t-1}$ to estimate importance of each of the $p$ covariates which are normalized into weights $\bv{\alpha} := \bracks{\alpha_1, \ldots, \alpha_p}^\top$ where $\alpha_1 + \ldots + \alpha_p = 1$ (line 5 of Algorithm~\ref{alg:matching}). There are many means of defining importance scores of covariates. A naive procedure computes $R^2_j$ for every univariate least squares regression of the historical responses on the historical values of the $j$th covariate and then normalizes by their sum, i.e. $\alpha_j := R^2_j / \sum_{k=1}^p R^2_k$. The $R^2$ metric is a unitless metric of explanatory importance to the response and hence a reasonable determinant of covariate importance (absolute linear coefficient estimates after standardization would be a similar approach). This procedure, although naive, provides good performance (when employed in simulations we will present in Section~\ref{sec:simulations}). However, it double-counts when covariates are collinear and we discuss a second-order improvement in this setting in Section~\ref{subsec:collinear_covariates}.

Matching subjects requires specification of a distance function that gauges proximity of a new entrant subject $\x_t$ and a reservoir subject $\x_r$ using $\bv{\alpha}$. There are many such distance functions that can be employed. We choose a covariate-weighted Euclidean distance metric,

\bneqn\label{eq:distance_metric_t}
d_t(\x_t, \x_r; \bv{\alpha}) := \sum_{j=1}^p \alpha_j (x_{t,j} - x_{r, j})^2
\eneqn

\noindent which is indexed by entrant $t$ because $\bv{\alpha}$ is recomputed at each entrant. We can contrast this to KK14 who employ a scaled squared Mahalanobis distance metric (a metric standard in the matching literature \citealp[e.g.][]{Greevy2004}),

\bneqn\label{eq:mahalanobis_distance}
d_M(\x_t, \x_r; \bv{S}) := \frac{t-p}{2p(t-1)}  (\x_t - \x_r)^\top \Sinv (\x_t - \x_r)
\eneqn

\noindent where $\bv{S}$ is the sample variance-covariance matrix using all $\x_1, \ldots, \x_t$ observations. The Mahalanobis distance adjusts for covariate collinearity but is not privy to how the covariates affect the response. Using the response information to weight covariates is the main source of our performance gain.

\label{paragraph:naive} Using Equation~\ref{eq:distance_metric_t}, we calculate the distance of the new entrant $\x_t$ to every subject in the reservoir and then locate the minimum distance reservoir subject (lines 6-7 of Algorithm~\ref{alg:matching}). To make the decision of whether or not to match the new entrant with the minimum distance reservoir subject, we need a threshold of the largest allowable distance $d_{t_\lambda}$. KK14 approximated their squared Mahalanobis distance threshold as the quantile of the appropriate scaled $F$ distribution when the covariate inputs are normally distributed. This likely works well for unimodal mostly symmetric continuous distributions but poorly otherwise (and especially poorly for discrete covariates).

For our distance function, there is no principled parametric assumption that would support using a distributional quantile. We use a resampling bootstrap procedure to estimate this threshold. For each resampling, we draw two unique subjects from $\x_1, \ldots, \x_t$ and compute $d_t$ (Equation~\ref{eq:distance_metric_t}) using the latest weights. With a very large number of resamplings, we will have a sampling distribution that is asymptotically convergent to the true distribution of $d_t$ over the data generating process that produces covariates and responses. We then designate $d_{t_\lambda}$ to be the $\lambda$ empirical quantile of the many resampling values (line 8 of Algorithm~\ref{alg:matching}). %This procedure is detailed in Algorithm~\ref{alg:resample_dt}.

If the distance between the new entrant and the minimum distance reservoir subject is less than $d_{t_\lambda}$, we match the two together and remove the minimum distance reservoir subject from the reservoir (lines 9-12 of Algorithm~\ref{alg:matching}). If the distance exceeds the critical threshold, the new entrant is randomized as in a Bernoulli trial and placed into the reservoir (lines 13-14 of Algorithm~\ref{alg:matching}).

\begin{algorithm}[t]
\caption{Our CARA sequential matching algorithm for subjects entering the experiment. The algorithm requires $t_0$ and $\lambda$ to be prespecified, which controls when we start creating matches and the ease of creating matches respectively.}
\begin{algorithmic}[1]
\For{$t \gets \braces{1,\ldots,n}$} \Comment{$n$ is the total sample size, fixed a priori}
   \If{$t \leq t_0$ \textbf{or} reservoir empty}
	\State $w_t \gets \bernoulli{1/2}$ and $\x_t$ is added to the reservoir \Comment{randomize}
    \Else
	\State $\bv{\alpha}$ weights are computed via the naive or stepwise procedure (see text)  %using $\x_1, \ldots, \x_{t-1}, y_1, \ldots, y_{t-1}$
%	\ForAll{$\x_r$ in the reservoir}
%		\State $d_t(\x_t, \x_r)$ is computed
%	\EndFor
\Statex  \hspace{1.2cm}\Comment{Find the minimum distance between the new subject and all reservoir subjects as well as index of the minimum distance in the reservoir}
	\State $d_{t_{r^*}} \gets \displaystyle\min_{r}\braces{d_t(\x_t, \x_r; \bv{\alpha})}$ 
	\State $r^* \gets \displaystyle \argmin_{r }\braces{d_t(\x_t, \x_r; \bv{\alpha})}$ \Comment{if ties exist, arbitrate via uniform draw}
	\State $d_{t_\lambda}$ is computed \Comment{Via the resampling procedure}
	\If{$d_{t_{r^*}} \leq d_{t_\lambda}$}  \Comment{The new subject is matched}
		\State $w_t \gets 1- w_{r^*}$ \Comment{assign the new subject the opposite assignment of subject $r^*$}
		\State Remove $\x_{r^*}$ from the reservoir
		\State Record $\angbrace{\x_t, \x_{r^*}}$ as a new match
	\Else \Comment{The new subject is not matched}
		\State $w_t \gets \bernoulli{1/2}$ and  $\x_t$ is added to the reservoir \Comment{randomize}
	\EndIf
    \EndIf
\EndFor
\end{algorithmic}
\label{alg:matching}
\end{algorithm}

A possible concern in all CARA designs is the danger of the practitioner doing multiple looks at the responses before the inferential analysis (the subject of Section~\ref{subsec:estimator}). This is not a concern herein as the practitioner does not actually see the responses (and in practice should be blinded to the responses). Only the design algorithm \qu{sees} the responses which are used exclusively to determine covariate weights and these weights are then in turn used to decide matches. Once the match structure (i.e. the set of subject indicies who are matched and the set of subject indicies remaining in the reservoir) is created, the response values themselves are independent of the allocations. This independence is due to the algorithm randomizing each matched pair to T-C or C-T with probability 50\% and employing a Bernoulli design to subjects that are placed in the reservoir.

\subsubsection{Improving Weighting in the Case of Collinear Covariates}\label{subsec:collinear_covariates}

In the setting of collinear covariates, we can implement a second-order improvement over the naive weight computation described in the previous section. The problem in the collinear covariate setting is that the naive univariate regressions create weights that double-count dimensions that covary (precisely what the Mahalanobis metric of Equation~\ref{eq:mahalanobis_distance} solves but our Equation~\ref{eq:distance_metric_t} does not address). This double-counting will yield subpar matches and thus, subpar estimation and subpar power. 

An intuitive approach that addresses this more general setting of collinear covariates would be to employ principle components analysis (PCA) to iteratively transform the covariate space and then assign weights based on the $R^2$ metric of each rotated orthogonal dimension. Equation~\ref{eq:distance_metric_t} would then be updated to compute distance based on differences in the $p$ principle components. Thus, ultimately the matches would be based on distance in this transformed space. However, the overarching goal in matching is to create close pairs in the original raw covariate space among covariates that strongly affect the response. Subjects who are proximal in the rotated principle components space may be very different in the original raw covariate space and hence dissimilar in response behavior. For this reason, this initial idea to use PCA to compute differential weights performed worse than the naive weighting procedure (in unshown simulation).

An alternative solution to obviate the double-counting in covariate weight estimation due to multicollinearity retains the original covariate space while employing a stepwise procedure as follows. For every subject entry $t$, we standardize each of the $p$ covariates and the responses of the previous subjects. We then subtract the current estimated additive treatment effect from the $y_i$'s when $w_i = 1$. Without this linear adjustment of the average treatment effect, the covariate imbalances for small values of $t$ will add variance in the estimation of the $\alpha_j$ weights. (Although it would be reasonable to adjust all covariates for the current allocation vector, we do not do so because this would invalidate the randomization test we develop in Section~\ref{subsubsec:permutation_test}). Then we use standard forward stepwise regression and at each step we record the squared partial correlations. After the $p$ steps, we normalize the squared partial correlations to become the weights. It is important to reiterate that this stepwise procedure is not being employed for model selection nor statistical inference, but merely to improve match quality to a second-order.

There are other alternative solutions such as a penalized multivariate regression where the weights are defined as the absolute coefficient estimates. A lasso L1 penalization would retain only the most salient covariates in each set of collinear covariates. However, the L1 shrinkage is harsh and retains very few or possible none of the covariates in low $n$ and low $R^2$ settings, defaulting the allocation to complete randomization. A ridge L2 penalization would likely perform better. Both penalized approaches would require a cross-validated grid search to select the regularization hyperparameter at each $t$. In addition to being computationally burdensome, the hyperparameter is selected by best response model fit, which is not the appropriate metric for covariate importance estimation.

All second order solutions that attempt to improve the weighting scheme under collinear covariates will be ad-hoc since the optimal procedure would require knowledge of the distributions of the covariates and knowledge of the explicit response model.

\subsection{Estimation and Hypothesis Testing}\label{subsec:estimator}

This section closely follows KK14. We consider inference where (a) with respect to covariates we do not wish to model their effect and (b) with respect to covariates we wish to model their effect assuming they combine linearly which we abbreviate NC and LC respectively. Considering both the population sampling model (Equation~\ref{eq:response_model}) abbreviated PO and the Neyman randomization model (Equation~\ref{eq:response_model2}) abbreviated NR, we consider a total of four inferential settings, NC/PO, NC/NR, LC/PO and LC/NR.

The NC vs LC settings distinguishes two different estimators and the PO vs NR settings distinguishes two types of testing. For estimating $\beta_T$ in the covariates-ignored case NC, we use a paired difference estimator for the matched pairs combined with the classic difference-in-means estimator detailed in Section~\ref{subsubsec:classic_test}. In the linear covariates case LC, we use an analogous OLS paired difference estimator combined with a classic OLS estimator detailed in section \ref{subsubsec:ols_test}. For testing, in the population model PO we test the population hypotheses $H_0: \betaT = \beta_0$ versus $H_a: \betaT \neq \beta_0$ detailed in Sections~ \ref{subsubsec:classic_test}~and~\ref{subsubsec:ols_test}. For the randomization model NR, we test Fisher's sharp null hypothesis using two similar randomization tests detailed in Section~\ref{subsubsec:permutation_test}.

%In our development of estimators and testing procedures, we always assume conditioning on $\mathcal{F}$, thus this notation is withheld going forward.

\subsubsection{Estimation and Testing for NC/PO}\label{subsubsec:classic_test}

%Upon completion of the experiment, there are $m$ matched pairs and $n_R$ subjects in the reservoir so that $2m + n_R = n$. We define $\Dbar$ as the estimator for the average of the differences of the $m$ matched pairs (treatment response minus control response) and $\YbarRT$, $\YbarRC$ as the estimators for the averages of the treatments and controls in the reservoir. We combine the estimators using a weight parameter, $B_T := w \Dbar + (1 - w) \parens{\YbarRTMinusYbarRC}$. We can find $w$ to minimize variance to obtain:
Upon completion of the experiment, there are $m$ matched pairs and $n_R$ subjects that were never matched and thus remaining in the reservoir (so that $2m + n_R = n$). We combine the estimator from the matched pairs and the estimator from the reservoir subjects linearly. The estimator in the matched pairs is the average pair differences (treatment response minus control response) denoted $\Dbar$. The estimator in the reservoir is $\YbarRT-\YbarRC$ where $\YbarRT$  and $\YbarRC$ are the means of the treated and control subjects in the reservoir respectively. Of course in the case where there are no matched pairs, we default to $\YbarRTMinusYbarRC$ exclusively and analogously when there are fewer than two treatments or controls in the reservoir, we default to $\Dbar$ exclusively. To obtain an efficient overall estimator, we take a convex combination of these two estimators where the mixture constant is chosen to minimize the variance,

\bneqn\label{eq:estimator}
\hat{\theta}_T = \frac{\sigsqR\Dbar + \sigsqDbar\parens{\YbarRTMinusYbarRC}}{\sigsqR + \sigsqDbar}.
%\var{B_T} &=& \frac{\sigsqR\sigsqDbar}{\sigsqR  + \sigsqDbar}.
\eneqn

%$B_T$ is unbiased because $\Dbar$ and $\YbarRTMinusYbarRC$ are unbiased. Standardizing $B_T$ gives a standard normal due to the assumption of normal noise. To create a usable test statistic, note that the true variances are unknown, so we plugin $\SsqDbar$, the matched pairs sample variance estimator, and $\SsqR$, the pooled two-sample reservoir variance estimator:

This estimator is always unbiased as the two constituent estimators are always unbiased (e.g. even in the case of misspecified covariate-treatment interactions). Under the population sampling model (Equation~\ref{eq:response_model}), $\hat{\theta}_T$ is distributed normally. As we do not know the variances, the estimator we employ is obtained in the conventional way by replacing the variances in Equation~\ref{eq:estimator} with their corresponding unbiased estimators,

\beqn
\SsqDbar &=& \oneover{m(m-1)}\sum_{i=1}^m \squared{D_i - \Dbar}, \\
\SsqR &=& \frac{\sum_{i=1}^{\nrt} \squared{Y_{R,T,i} - \Ybar_{R,T}} + \sum_{i=1}^{\nrc} \squared{Y_{R,C,i} - \Ybar_{R,C}}}{\nr -2} \parens{\oneover{\nrt} + \oneover{\nrc}}
\eeqn

\noindent where  $\nrt$ and $\nrc$ denote the number of subjects in the treated and control groups in the reservoir respectively. Even though technically $\nrt$ and $\nrc$ are random variables, we ignore this source of variation as is typically done in the randomized trial literature. Applying these expressions, our estimator becomes

\bneqn\label{eq:our_estimator}
B_T = \frac{\SsqR \Dbar + \SsqDbar \parens{\YbarRTMinusYbarRC}}{\SsqR + \SsqDbar}.
\eneqn

%We use the notation $\nrt$ and $\nrc$ to be the number of treatments and controls in the reservoir ($\nr = \nrt+\nrc$). In practice, $\nrt$ is a random quantity, a binomial distribution with size $\nr$ and probability of one half. However, we will assume $\nrt$ and $\nrc$ are fixed as an approximation. A more careful calculation could include the randomness of $\nrt$ and $\nrc$.
To create a statistic to test the null hypothesis of a specific treatment effect $\beta_0$, we subtract $\beta_0$ and divide by the standard deviation. Since the variance of $\hat{\theta}_T$ is $\parens{\sigsqR\sigsqDbar} / \parens{\sigsqR  + \sigsqDbar}$ and we have $\SsqDbar \convp \sigsqDbar$, $\SsqR \convp \sigsqR$, a Wald test follows since

\bneqn\label{eq:clt_true_var}
\sqrt{\frac{\SsqR + \SsqDbar}{\SsqR\SsqDbar}} \parens{B_T - \beta_0} ~\convd~ \stdnormnot
\eneqn

\noindent and a Wald interval with endpoints $B_T \pm z_0 \times S_R S_{\Dbar} / (\SsqR + \SsqDbar)^{1/2}$ where $z_0$ is the standard normal quantile appropriate for the level of coverage also follows.

%When is this estimator more efficient than the standard classic estimator, $\Delta\Ybar := \YbarT - \YbarC$? In other words, when is $\sigsqDeltaYbar / \sigsqDbar > 1$? Assuming perfect balance in its treatment allocation ($n_T = n_C = \overtwo{n}$) for the classic estimator and taking the expectation over both noise and treatment allocation, it can be shown that the variances are:

Our estimator $B_T$ will be more accurate than the naive estimator $\YbarT - \YbarC$ when  $\sigsq_{\YbarT - \YbarC} / \sigsqDbar > 1$. This is easiest to explore when $n_T = n_C = \overtwo{n}$. In this case it follows that 

%\bneqn\label{eq:var_sigsq_Dbar}
\beqn
\sigsqDbar = \frac{1}{m^2}  \sum_{k=1}^m \squared{z_{T,k} - z_{C,k}} + \frac{2}{m} \sigsq_e, ~~\sigsq_{\YbarT - \YbarC} \approx \frac{4}{n^2} \sum_{i=1}^{n} z_{i}^2 + \frac{4}{n}\sigsqe.
\eeqn
%\eneqn

%This means that the better the matching, the smaller $\sum_{k=1}^m \squared{z_{T,k} - z_{C,k}}$ will be, the smaller the variance becomes, and the higher the power. If we further allow $n_R = 0$ (\textit{all} the subjects matched), then it's clear that $\sigsqDeltaYbar / \sigsqDbar > 1$ if and only if $\sum_{k=1}^m z_{T,k} z_{C,k} > 0$. Note that the approximation in the last expression is due to ignoring covariance terms which do not exist when conditioning on $n_T$ and $n_C$. %\footnote{This works even if the observations in the reservoir have wildly different $z_i$'s since they'll be subsumed in the $s^2_R$ calculation. The weighting scheme in equation \ref{eq:estimator} will then down-weight the reservoir's influence.}

\noindent Since the closer units are matched, the smaller the $\squared{z_{T,k} - z_{C,k}}$ terms become, the more accurate the estimation and the higher the power. In the special case that all units are matched, the condition $\sigsq_{\YbarT - \YbarC} / \sigsqDbar > 1$ holds when $\sum_{k=1}^m z_{T,k} z_{C,k} > 0$.

%It is well known that , thus plugging in the sample variances will preserve the normal convergence in equation \ref{eq:clt_true_var} by Slutsky's theorem, thereby providing us with a usable $Z$-like test statistic.

%Although we have normal convergence for this test statistic, we do not have an idea about how quickly it converges.  Thus, it makes sense to develop a complementary non-parametric test.

\subsubsection{Estimation and Testing for LC/PO}\label{subsubsec:ols_test}

We can also construct a linearly-combined match-reservoir estimator that accounts for the effect of the covariates by employing linear regression. This requires providing analogue OLS estimators for both $\Dbar$ and $\YbarRTMinusYbarRC$.

We consider the following model for the $k = 1, \ldots, m$ matched pairs: 

\beqn
D_k = \beta_{0,D} + \beta_{1,D} \Delta x_{1,k} + \ldots + \beta_{p,D} \Delta x_{p,k} +  \epsilonrv_{k,D}, ~~\epsilonrv_{k,D} \iid \normnot{0}{\tausq_D}
\eeqn

\noindent where $\Delta x_{1,k}, \ldots, \Delta x_{p,k}$ denote the differences between treatment and control within match $k$ for each of the $p$ covariates. The OLS estimator for the intercept $\beta_{0,D}$ which we denote $B_{0,D}$ is the analogue of $\Dbar$.

Similarly, for the responses $i = 1, \ldots, n_R$ in the reservoir, we consider the model:

\beqn
Y_i = \beta_{0,R}  + \beta_{T,R} w_i + \beta_{1,R} x_{1,i} + \ldots + \beta_{p,R} x_{p,i} +  \epsilonrv_{i,R}, ~~\epsilonrv_{i,R} \iid \normnot{0}{\tausq_R}.
\eeqn

The OLS estimator for the intercept $\beta_{0,R}$ which we denote $B_{0,R}$ is the analogue of $\YbarRTMinusYbarRC$. Combining the two estimators, we arrive at the estimator analogous to Equation~\ref{eq:our_estimator},

\bneqn\label{eq:our_ols_estimator}
B_T^{OLS} = \dfrac{S^2_{B_{T,R}} B_{0,D} + S^2_{B_{0,D}} B_{T,R}}{S^2_{B_{T,R}} + S^2_{B_{0,D}}}
\eneqn

\noindent where $S^2_{B_{T,R}}$ is the sample variance of $B_{T,R}$ and $S^2_{B_{0,D}}$ is the sample variance of $B_{0,D}$. We can construct a Wald test due to the fact that

\bneqn\label{eq:clt_true_var_linear} %\frac{B_{T,OLS} - \beta_0}{\se{B_{T,OLS}}} \approx
%\beqn
\sqrt{\frac{S^2_{B_{T,R}} + S^2_{B_{0,D}}}{S^2_{B_{T,R}} S^2_{B_{0,D}}}} \parens{B_T^{OLS} - \beta_0} ~\convd~ \stdnormnot
\eneqn
%\eeqn

\noindent analogous to Equation~\ref{eq:clt_true_var} along with the analogous Wald interval with endpoints $B_T^{OLS} \pm z_0 \times S_{B_{T,R}} S_{B_{0,D}} / (S^2_{B_{T,R}} + S^2_{B_{0,D}})^{1/2}$.

\subsubsection{Estimation and Testing for NR}\label{subsubsec:permutation_test}

We present two randomization tests for the sharp null (also called \qu{Fisher's exact tests} as in \citealt[Chapter 5]{Imbens2015}). We draw many null treatment-control assignments vectors $\w$ and for each, compute an estimate of the treatment effect. We use these estimates to construct a null estimator distribution and then assess the quantile of the estimate computed from the actual experimental assignment. Various functions of the data can be employed as long as they sensibly capture the effect of the treatment \citep[see discussion in][Chapter 5.5]{Imbens2015}. We employ the two described previously: our classic-combined estimate (whose estimator is $B_T$ of Equation~\ref{eq:our_estimator}) and our OLS-inspired estimate (whose estimator is $B_T^{OLS}$ of Equation~\ref{eq:our_ols_estimator}).

How can we randomly produce the null $\w$ vectors? These assignments must be randomized in the same fashion that was employed for the actual experimental assignments as \qu{one should analyze as one designs} \citep[p. 105]{Rosenberger2016} and \qu{[one should keep] only simulated randomizations that satisfy the balance criterion} i.e. in the case of rerandomization designs \citep[p. 1266]{Morgan2012}.

We first describe why under different null $\w$ vectors, the reservoir subjects and matches remain exactly the same. Under the Neyman randomization model and the sharp null, Equation~\ref{eq:response_model2} becomes $Y_t = f(\x_t)$. Since the covariates are fixed and the order of subject entry is kept constant, the response values are fixed for all $\w$. Thus the computation of the $\bv{\alpha}$ weights via the naive or stepwise procedures (line 5 of Algorithm~\ref{alg:matching}) remain the same across all $\w$ as their values result from a deterministic OLS regression of fixed response values on fixed covariate values. Thus all distances computed between the entering subject $\x_t$ and the reservoir subjects $\x_r$ are the same. Keeping the resamplings constant, the computation of $d_{t_\lambda}$ (line 8 of Algorithm~\ref{alg:matching}) is then independent of $\w$. Thus both the indices of the matched pairs and the indices of the subjects left in the reservoir will be independent of $\w$.

Given the structure of reservoir subjects and matches pairs, we can draw null $\w$ vectors as follows. For subjects in a matched pair, assign 1-0 or 0-1 with equal probability. For the subjects in the reservoir, first record the number in each arm in the actual experimental run ($\nrt$ and $\nrc$). Then draw a reservoir assignment by a shuffling of a $n_R$-tuple with $\nrt$ elements equal to 1 and $\nrc$ elements equal to 0. This resultant testing procedure is a \qu{conditional randomization test} \citep[page 104, Equation 6.6]{Rosenberger2016}. Each of these null vectors have equal probability of being chosen as the actual experimental run assignment. Since we cannot possibly analyze the exponential number of possible assignments, $2^m \times \binom{n_R}{\nrt}$ (which e.g. is about 220 million if $n=50$ and there are $m=20$ matches and $\nrt = 4$ of the $n_R = 10$), we sample a large number of null assignments and thus use a \qu{Monte-Carlo randomization test} which is accurate to any desired margin of error \citep[Chapter 6.9]{Rosenberger2016}.

Since tests and confidence intervals share a duality and both the NC and LC models assume an additive
treatment effect, a confidence interval can be produced from both of these randomization tests as the set of all values for which the actual experimental estimate would not reject a hypothesized shift in the sharp null. There are efficient algorithms for such a computation \citep[see e.g.][]{Garthwaite1996}.

%we need to permute the assignment to $T$ and $C$ in an informed way. In practice we create Monte-Carlo samples of such assignments and for each sample obtain the behavior of the test statistic. This is done for both the classical and covariate adjusted estimators. The permutation distribution for the matched pairs is obtained by flipping a coin for each matched pair in the labling of the items, $(T,C)$ or $(C,T)$. In the case of the items in the reservoir, we permute the labels of the assignments. Note that this does not alter $n_{R,T}$ or $n_{R,C}$

\subsection{Choosing the Hyperparameter Values}\label{subsec:hyperparameters}

As $t_0$ increases, the quality of the matches improves but there are fewer matches. Also, as $\lambda$ decreases, the quality of the matches improves but there are fewer matches. The optimal values of these two design hyperparameters are unknown because they depend on the relationship between the response and covariates and choice of estimator. In KK14, we set $\lambda = 5\%$ just like the default size in hypothesis testing as this was an intuitive choice. Also in KK14, we set $t_0 = p$ (i.e. begin matching immediately when the sample variance-covariance matrix of the covariates is invertible). We have since learned that increasing $t_0$ to a large proportion of total sample size (we default to 35\%$n$) and increasing $\lambda$ modestly (we default to 10\%) leads to much better estimator performance due to the increase in the number of matches with negligible cost in match quality. %For example, if $n=100$, $p=1$, $\lambda = 5\%$ and $t_0 = 35$, then the average reservoir size is 22.25. If $t_0$ increases to 50, then the average reservoir size increases to 28.71. At $\lambda = 10\%$, these the reservoir size averages are 11.14 and 17.49 respectively.

%Using the stylized model from KK14 it follows that the number in the reservoir grows as the number of burn in increases and lambda decreases.
%
%Even though the number of matches decreases in these parameters the quality of the matches improves. The choice of these parameters depends
%
%on the relationship between the response and covariates and choice of estimator. In KK14 it was determined the number in the reservoir with small
%
%burn-in is 10 when lamda=.10 and 20 when lambda=.05. These values do not increase much with increase in the burn in. As the burn in ranges from 30 to 50
%
%for n=100 the average number in the reservoir increases from 21.25 to 28.71 when lambda=.05 and 10.48 to 17.49 when lambda=.10. Although, arbitrary we have found
%
%that having a 35\% burn in with lambda=.10 performs well with little increase in the number in the reservoir for the gain one gets in the quality of the matches.

\section{Results}\label{sec:results}

\subsection{Simulated Data}\label{sec:simulations}

We first demonstrate the performance of our CARA design using simulated data. In this section, we simulate experimental subject data under the response model of independent $Y_i = \beta_1 x_{i,1} + \beta_2 x_{i,2} + \beta_3 x_{i,1}^2+ \beta_T w_i + \epsilonrv_i$ where the two covariates $x_{i,1}$ and $x_{i,2}$ are drawn from $\normnot{1}{1}$ distributions and the error terms $\epsilonrv_i$'s are drawn from an iid $\normnot{0}{1}$ distribution. This model and error variance were chosen to induce adequate separation over the power we measure in the different simulation settings. If the subjects are well-matched, then the power should be very high as over the many simulations, we compare two Gaussian response distributions with variance of $1/n$ differing only by a small shift in mean; if the subjects are not well-matched, the variance is much higher and thus power would be considerably lower. %This response function is non-linear (there is a quadratic term for the first covariate) and collinear. 

We have a number of simulation settings, the first being (A1: treatment effect) $\beta_T = 1$ and (A2: no treatment effect) $\beta_T = 0$ where the latter setting was simulated to assess whether the simulated tests are correctly sized. We next considered sample sizes of (N1: 50) and (N2: 100). We also varied the dependence of the two covariates (X1: uncorrelated) where the two covariates are independent and (X2: correlated) where the correlation between the two covariates is .75. The latter setting was simulated to assess designs under collinearity. The variable importance was also varied (B1: uniform), where all $\beta_j = 1$ and (B2: varied), where $\beta_1 = 6, \beta_2 = 1$ and $\beta_3 = 2$. The latter (B2) was simulated to assess a response model with covariates with different relative importance. 

We then allocated $\w$ under a variety of sequential experiment design procedures: (D1: Bernoulli) trial, (D2: BCRD), (D3: Biased) Coin of \citet{Efron1971} with his recommended coin bias of 2/3, (D4: Minimization) of \citet{Pocock1975}, (D5: Atkinson's) biased coin explained in Section~\ref{sec:introduction}, (D6: KK14) sequential matching without differentially weighting the covariates, (D7: naive) sequential covariate weighted algorithm where the weights are the univariate $R^2$'s normalized and the (D8: stepwise) sequential weighted algorithm. For (D6 - D8) we set $\lambda = 10\%$ and $t_0$ to be 35\% of the sample size $n$. For (D7) and (D8) we resample 500 subject pairs to compute the matching threshold distance, $d_{t_\lambda}$. 

Upon completion of the sequential experiment we computed different estimators: the (E1: classic) $\YbarT - \YbarC$, (E2: OLS) after a regression of the responses on an intercept, the three covariates and $\w$, we employed the slope coefficient for the $\w$ term. Both (E1) and (E2) were recorded for designs (D1 - D5). For the sequential matching designs (D6 - D8), we analogously computed the (E3: combined classic) estimate $B_T$ of Equation~\ref{eq:our_estimator} that uses information from both matches and the reservoir and the (E4: combined OLS) estimate $B_T^{OLS}$ of Equation~\ref{eq:our_ols_estimator} that also uses information from both matches and the reservoir. 

To test appropriately under both the population model (Equation~\ref{eq:response_model}, where the null hypothesis is the zero population mean effect of treatment) and the Neyman Randomization model (Equation~\ref{eq:response_model2}, where the only randomness comes from $\w$ and the sharp null is considered), we use the (T1: NBT) the normal theory-based test which is an asymptotically valid $Z$ test for all (D1 - D5) and Wald's asymptotically valid Z-test of Equations~\ref{eq:clt_true_var} and \ref{eq:clt_true_var_linear} for (D6 - D8) and the (T2: RAN) which denotes the randomization test with one exception. We omit an exact test for (D4), the design of \citet{Pocock1975} as it would be computationally cumbersome (and we are not aware of such an algorithm appearing in the literature). For all randomization tests, we draw 501 null vectors (one more than 500 to avoid ambiguity at the exact $\alpha$ quantile).

There are a total of $2 (N) \times 2 (A) \times 2 (X) \times 2 (B) \times 8 (D) \times 2 (E) \times 2 (T)$ less the 32 settings for the omission of (T2) in (D4) for a total of 480 simulation settings. In each setting, we run 10,000 duplicates and for each duplicate, we record the estimate of the treatment effect and whether or not the null hypothesis is rejected at significance level 5\%. Over the duplicates, we compute the mean squared error and the power (or size) for each simulation setting. Table~\ref{tab:power50} displays the power estimates for $n = 50$ over the 120 = 480 / 2~(N)~/ 2 (A) settings. Power estimates for $n = 100$ can be found in Table~\ref{tab:power100} in the Supporting Information. We choose to illustrate the smaller sample size because power at the higher sample size was near 1 for many settings, making design comparisons difficult.

\begin{table}[htp]
%\small
\centering
\begin{tabular}{ccc|cc|cc}
																		&																					&				& \multicolumn{4}{c}{Betas} 																																	\\ 
																		&																					&				& \multicolumn{2}{c|}{B1: Uniform} & \multicolumn{2}{c}{B2: Varied}																\\ \cline{4-7}
																		&																					&				& \multicolumn{2}{c|}{Covariates } & \multicolumn{2}{c}{Covariates}\\
																		&																					&				& \multicolumn{2}{c|}{ Correlated?} & \multicolumn{2}{c}{Correlated?}\\
  																		&   																				&  			 & No				& Yes	 		& No				& Yes	 																												\\ \cline{4-7}
Design															&	Estimate																&	Test	 & \multicolumn{4}{c}{Power} 																														\\  \hline
 \multirow[c]{4}{*}{D1: Bernoulli}  		& \multirow[c]{2}{*}{E1: Classic} 										&	 NBT 		& .164 			& .135 		& .068 			& .062\\
 																		& 												& RAN  		& .170 			& .140 		& .061 			& .059 \\ \cline{2-7}
 																		& \multirow[c]{2}{*}{E2: OLS}	 			&	 NBT 		& .532 			& .539 		& .243 			& .239 \\ 
 																		& 												& RAN  		& .543 			& .539		& .258			& .251\\ \cline{1-7}
 \multirow[c]{4}{*}{D2: BCRD}  		& \multirow[c]{2}{*}{E1: Classic} 											&	 NBT 		& .171 			& .137 		& .064 			& .061\\
 																		& 												& RAN  		& .171 			& .141 		& .067 			& .064 \\ \cline{2-7}
 																		& \multirow[c]{2}{*}{E2: OLS}				&	 NBT 		& .542 			& .546 		& .250 			& .237 \\ 
 																		& 												& RAN 			 & .548 		& .549		& .257			& .250\\ \cline{1-7}
 \multirow[c]{4}{*}{D3: Biased} 				& \multirow[c]{2}{*}{E1: Classic} 									&	 NBT 		& .163 			& .134 		& .059 			& .054 \\ 
 																		& 												& RAN  		& .174 			& .135		& .058			& .064\\ \cline{2-7}
 																		& \multirow[c]{2}{*}{E2: OLS}				&	 NBT 		& .543			& .542 		& .238 			& .242 \\ 
 																		& 												& RAN  		& .549 			& .542 		& .254 			& .257\\ \cline{1-7}
 \multirow[c]{4}{*}{D4: Minimization} & \multirow[c]{2}{*}{E1: Classic} 											&	 NBT 		& .123 			& .080		& .022 			& .013 \\ 
 																		& 												& RAN  		& -- 			& -- 		& -- 			& --\\ \cline{2-7} 
 																		& \multirow[c]{2}{*}{E2: OLS}	 			&	 NBT		& .567 			& .555 		& .226 			& .224 \\ 
 																		& 												& RAN  		& -- 			& -- 		& --			& --\\ \cline{1-7} 
 \multirow[c]{4}{*}{D5: Atkinson} 				& \multirow[c]{2}{*}{E1: Classic} 								&	 NBT 		& .059 			& .030 		& .001 			& .000 \\ 
 																		& 												& RAN 			 & .356 		& .312		& .106			& .096\\ \cline{2-7}
 																		& \multirow[c]{2}{*}{E2: OLS}				&	 NBT 		& .548			& .553 		& .253 			& .261 \\ 
 																		& 												& RAN  		& .554 			& .551 		& .262 			& .270\\ \cline{1-7}
 \multirow[c]{4}{*}{D6: KK14} 				& \multirow[c]{2}{*}{E3: Combined Classic} 							&	 NBT 		& .718 			& .667 		& .276 			& .264 \\ 
 																		& 												& RAN  		& .668 			& .613 		& .248 			& .218\\ \cline{2-7}  
 																		& \multirow[c]{2}{*}{E4: Combined OLS}	 &	 NBT 		& .803 			& .804 		& .680 			& .678 \\ 
																		& 												& RAN  		& .731 			& .730 		& .614 			& .610\\ \cline{1-7} 
 \multirow[c]{4}{*}{D7: naive} 				& \multirow[c]{2}{*}{E3: Combined Classic} 							&	 NBT 		& .766 			& .763 		& .543 			& .389 \\ 
 																		& 												& RAN  		& .711 			& .727 		& .488 			& .349\\ \cline{2-7}   
 																		& \multirow[c]{2}{*}{E4: Combined OLS}	 &	 NBT 		& .826 			& .817 		& .803 			& .736 \\ 
 																		& 												& RAN  		& .765 			& .754 		& .751			& .688\\ \cline{1-7}  
 \multirow[c]{4}{*}{D8: stepwise} 			& \multirow[c]{2}{*}{E3: Combined Classic} 							&	 NBT 		& .772 			& .789 		& .577 			& .642 \\ 
	 																	& 												& RAN  		& .724 			& .746		& .528			& .589\\ \cline{2-7}    
 																		& \multirow[c]{2}{*}{E4: Combined OLS}	 &	 NBT 		& .834 			& .834 		& .811 			& .824 \\ 
 																		& 												& RAN  		& .763 			& .781 		& .764 			& .775\\ \hline
\end{tabular}
\caption{Experimental power (A1) at $n=50$ under the 120 varied simulation settings rounded to the nearest tenth of a percent.}
\label{tab:power50}
\end{table}

It is inappropriate to compare powers for (B1) uniform variable importance and (B2) varied variable importance because they are different response models. Four comparisons should be made for each table column in Tables~\ref{tab:power50} (and \ref{tab:size50}) among the designs. These four comparisons are the distinct E/T pairs: Classic~/ NBT, Classic / RAN, OLS / NBT and OLS / RAN.

The (E2) OLS estimates perform better than the (E1) classic estimates across all (D1-D8, X1, X2, B1 and B2) settings. This is expected as the OLS is able to adjust for the linear component within the response model. The sequential designs of (D1-D3) neither use the covariate values nor responses and they all have similar performance within E/T pairs. For (D4-D5), these designs have the advantage of using the covariate values of the entrants. However, their (T1: NBT, E1: Classic) are beaten by (D1-D3). This is expected as (D4,D5) are heavily restricted designs and these designs \qu{[change] the distribution of the test statistic, most notably by decreasing the true standard error, thus traditional methods of analysis that do not take this into account will result in overly `conservative' inferences in the sense that tests will reject true null hypotheses less often than the nominal level...} \citep[page 1264]{Morgan2012}. This phenomenon does not seem to manifest in the OLS estimator (and does not affect the T2: RAN estimator whatsoever \citealp[Section 2.2]{Morgan2012}). Minimization's (D4) OLS estimator beats (D1-D3) for (B1) but not for (B2). The latter failure is likely due to blocking's inability to address collinearity. Atkinson's design (D5) which assigns by accounting for off-diagonal entries within $\inverse{\X^\top \X}$, is devised to handle collinearity well and beats (D1-D3) in (B2, E2). However, the advantage of (D4,D5) in (E2) OLS is marginal which is expected because \citet[Section 2.3.2]{Kapelner2020} shows that there are only nominal gains to be had by balancing covariates a priori if OLS is employed.

%Comparing the classic designs of (D1, D3) and (D4) was covered in KK14 and we largely see the same pattern repeat here under all (E / T) settings among both (B1) and (B2) which we briefly reiterate. The Biased Coin provides essentially the same power as the random Bernoulli design. Minimization is hypothesized to perform better for NBT than the Biased Coin as it balances the covariates between the two treatment arms but it does not (similar to the results of KK14). 

The KK14 design (D6) beats all classic designs (D1-D5) by a substantial amount in every E/T cell. This is due to the well-established performance boost of using matched pairs, performance that gets even better when the response model is non-linear \citep{Greevy2004}.

Let us now compare our two new proposed methods, (D7) naive on-the-fly covariate importance weighting and (D8) stepwise weighting to (D6) KK14 which weights all covariates equally. Under all (E1, E2, B1, B2, X1, X2) we observe that (D8) $>$ (D7) $>$ (D6). These power improvements are most substantial in the (B2/X2) setting as (D8) is especially designed for differential covariate importance where the covariates are collinear.

%. The power advantage of (D8) $>$ (D7) $>$ (D6) is most starkly apparent in (B2) and (X2), with correlated covariates. 

%Under RAN and (B1), our new methods (D5, D6) on average have higher power than the older (D4), with stepwise generally doing better than naive. Under all (B1, B2, X1, X2), the power advantage of (D6) over (D5) over (D4) is much stronger, under independent covariates (X1) it is $75\% > 70\% > 46\%$ and under correlated covariates (X2) it is $75\% > 63\% > 22\%$. The advantage of the stepwise (D6) over the naive (D5) on-the-fly weighting is most apparent in the case of correlated covariates (X2) because the estimated weights are more reflective of the true variable importances in the response model. 

%In summary, all matching designs (D4-D6) outperform the classical designs (D1-D3). The new weighting designs (D5, D6) works at least a well as KK14 (D4) when it is expected to confer little advantage, e.g. in the setting of relatively uniform variable importances. In the cases where there are different variable importances, the new weighting designs (D5, D6) confer a substantial advantage over KK14 (D4) and the advantage of stepwise weighting (D6) over the naive (D5) is most stark when there is dependence among the covariates.

Table~\ref{tab:size50} displays the sizes for the same settings at $n = 50$. Size estimates for $n = 100$ can be found in Table~\ref{tab:size100} in the Supporting Information. There is no qualitative difference between Table~\ref{tab:size50} and Table~\ref{tab:size100}.

\begin{table}[htp]
%\small
\centering
\begin{tabular}{ccc|cc|cc}
																		&																					&				& \multicolumn{4}{c}{Betas} 																																	\\ 
																		&																					&				& \multicolumn{2}{c|}{B1: Uniform} & \multicolumn{2}{c}{B2: Varied}																\\ \cline{4-7}
																		&																					&				& \multicolumn{2}{c|}{Covariates } & \multicolumn{2}{c}{Covariates}\\
																		&																					&				& \multicolumn{2}{c|}{ Correlated?} & \multicolumn{2}{c}{Correlated?}\\
  																		&   																				&  			 & No				& Yes	 		& No				& Yes	 																												\\ \cline{4-7}
Design															&	Estimate																&	Test	 & \multicolumn{4}{c}{Size} 																														\\  \hline
 \multirow[c]{4}{*}{D1: Bernoulli}  		& \multirow[c]{2}{*}{E1: Classic} 											&	 NBT 		& .0471 		& .0500 	& .0503 		& .0527\\
 																		& 													& RAN  		& .0520 		& .0544 	& .0534		& .0491 \\ \cline{2-7}
 																		& \multirow[c]{2}{*}{E2: OLS}	 				&	 NBT 		& .0461 		& .0532	& .0452		& .0492 \\ 
 																		& 													& RAN  		& .0521 		& .0548	& .0492		& .0527\\ \cline{1-7}
 \multirow[c]{4}{*}{D2: BCRD}  		& \multirow[c]{2}{*}{E1: Classic} 												&	 NBT 		& .0481 		& .0472	& .0491		& .0501\\
 																		& 													& RAN  		& .0525 		& .0558 	& .0502		& .0541 \\ \cline{2-7}
 																		& \multirow[c]{2}{*}{E2: OLS}	 				&	 NBT 		& .0473 		& .0491	& .0485		& .0470 \\ 
 																		& 													& RAN  		& .0523 		& .0506	& .0522		& .0484\\ \cline{1-7}
 \multirow[c]{4}{*}{D3: Biased} 				& \multirow[c]{2}{*}{E1: Classic} 										&	 NBT 		& .0524 		& .0535	& .0443 		& .0454 \\ 
 																		& 													& RAN  		& .0520 		& .0555	& .0546 		&.0556\\ \cline{2-7}
 																		& \multirow[c]{2}{*}{E2: OLS}	 				&	 NBT 		& .0493 		& .0467	& .0475		& .0483 \\ 
 																		& 													& RAN  		& .0523 		& .0531	& .0507		& .0514\\ \cline{1-7}
 \multirow[c]{4}{*}{D4: Minimization} & \multirow[c]{2}{*}{E1: Classic} 												&	 NBT 		& .0164* 		& .0084* 	& .0084* 		& .0057* \\ 
 																		& 													& RAN  		& -- 			& -- 		& -- 			& --\\ \cline{2-7} 
 																		& \multirow[c]{2}{*}{E2: OLS}	 				&	 NBT 		& .0318* 		& .0319* 	& .0222* 		& .0205* \\ 
 																		& 													& RAN  		& -- 			& -- 		& --			& --\\ \cline{1-7} 
 \multirow[c]{4}{*}{D5: Atkinson} 				& \multirow[c]{2}{*}{E1: Classic} 									&	 NBT 		& .0012* 		& .0535*	& .0003* 		& .0001* \\ 
 																		& 													& RAN  		& .0543 		& .0555	& .0524		&.0520\\ \cline{2-7}
 																		& \multirow[c]{2}{*}{E2: OLS}	 				&	 NBT 		& .0527 		& .0467	& .0514		& .0484 \\ 
 																		& 													& RAN  		& .0511 		& .0531	& .0500		& .0528\\ \cline{1-7}
 \multirow[c]{4}{*}{D6: KK14} 				& \multirow[c]{2}{*}{E3: Combined Classic} 								&	 NBT 		& .0671* 		& .0680*	& .0677* 		& .0683* \\ 
 																		& 													& RAN  		& .0527		& .0498	& .0537		& .0511\\ \cline{2-7}  
 																		& \multirow[c]{2}{*}{E4: Combined OLS}	 	&	 NBT 		& .0751* 		& .0707*	& .0688* 		& .0663* \\ 
																		& 													& RAN  		& .0498		& .0535	& .0546		& .0500\\ \cline{1-7} 
 \multirow[c]{4}{*}{D7: naive} 				& \multirow[c]{2}{*}{E3: Combined Classic} 								&	 NBT 		& .0678*		& .0685* 	& .0688* 		& .0662* \\ 
 																		& 													& RAN  		& .0473		& .0508 	& .0491		& .0478\\ \cline{2-7}   
 																		& \multirow[c]{2}{*}{E4: Combined OLS}	 	&	 NBT 		& .0715*		& .0743* 	& .0682* 		& .0728* \\ 
 																		& 													& RAN  		& .0482		& .0494	& .0545		& .0529\\ \cline{1-7}  
 \multirow[c]{4}{*}{D8: stepwise} 			& \multirow[c]{2}{*}{E3: Combined Classic} 								&	 NBT 		& .0690*		& .0723*	& .0661*		& .0648* \\ 
	 																	& 													& RAN  		& .0533		& .0530	& .0521 		& .0535\\ \cline{2-7}    
 																		& \multirow[c]{2}{*}{E4: Combined OLS}	 	&	 NBT 		& .0751* 		& .0709* 	& .0681* 		& .0662* \\ 
 																		& 													& RAN  		& .0515		& .0529	& .0496		& .0517\\ \hline
\end{tabular}
\caption{Experimental size at $n=50$ under the 120 varied simulation settings rounded to the nearest hundredth of a percent. Asterisks indicate the test is likely not properly sized as the null hypothesis that size equals the purported 5\% tested at $\alpha = 5\%$ with Bonferroni corrections for multiple hypothesis testing was rejected (the rejection is less than 0.04231 and greater than 0.05769).}
\label{tab:size50}
\end{table}

We observed previously in Table~\ref{tab:power50} that in on-the-fly matching designs of (D6-D8), the (T1: NBT) setting has higher power than the (T2: RAN) setting. Part of this reason may be because the Wald tests of Equation~\ref{eq:clt_true_var}, Equation~\ref{eq:clt_true_var_linear} and Equation~6 in KK14 are asymptotic and thus employing the normal quantiles as the critical cutoffs are anti-conservative. This intuition is confirmed by the size estimates in Table~\ref{tab:size50}; many of the (T1: NBT) settings for the matching designs (D6-D8) are statistically significantly different than the purported $\alpha = 5\%$ echoing Table~3 of KK14. The (T2: RAN) tests for all designs are correctly sized.

This simulation offers a clear practical recommendation; one should allocate using the new design (D8), estimate the treatment effect with (E2: OLS) and run the hypothesis test using (T2: RAN).

%We then assign treatments under a variety of sequential experimental designs and then test using four different testing procedures under size $5\%$. We also vary the sample size. Table~\ref{tab:simulation_settings} details all xxx settings  which we call simulation cells. 

%\begin{table}[htp]
%\begin{tabular}{l|ccccc}
%&          				& Covariate  										& Experimental    &   						& $H_0$ \\
%& Sample 			& distribution  								   & Assignment  		& Test  			  & Rejection \\
%& sizes (n) 			& covariances ($\bv{\Sigma}$)  & Designs 			& Statistics 		& Test \\ 
%Setting A & 50 & $\I_3$ & Bernoulli Trial & $\YbarT - \YbarC$ & Student's T-test \\
%Setting B & 100 & $\threebythreemat{1}{.75}{0}{.75}{1}{0}{0}{0}{1}$ & Efron's Biased Coin & OLS & Randomization Test\\
%Setting C & 200 & \\
%Setting D &  &   \\
%Setting E &  &   \\
%\end{tabular}
%\caption{Simulation settings.}
%\label{tab:simulation_settings}
%\end{table}

\subsection{No Investigator Bias due to Guessing}\label{sec:investigator_bias}

The main thrust of this paper is our CARA design's performance when estimating and testing the treatment effect. Another important issue in sequential experimental design is the degree to which an investigator can guess future allocations and thus nefariously recruit individuals at the opportune time in order to manipulate the experimental outcome. When considering investigator bias, different experimental designs are evaluated for different investigator \emph{guessing strategies}. A guessing strategy decides $w_t$ as a function of previous allocations, covariates and possibly responses. The strategy's \emph{guess rate} is the expected percentage of correct guesses over $t = 1, \ldots, n$.

Restricted designs are more guessable than less restricted designs. For example, the Bernoulli trial is unguessable whereby any guessing strategy would yield a guess rate of 50\%, the alternating design T, C, T, C, etc. with the same guessing strategy would have a trivial guess rate of 100\%. When employing BCRD, \citet{Blackwell1957} proves that the \qu{convergence guessing strategy}, whereby you guess $w_t$ deterministically to be the arm that has occurred least in $w_1, \ldots, w_{t-1}$, provides the highest possible guess rate which is $n/2 + 2^{n-1} / \binom{n}{n/2}$ which is for example 55.78\% when $n=100$.

We investigate the convergence guessing strategy for all designs of the previous section. We simulated $n=100$ subjects using (X2: correlated) covariates with (B1: uniform) variable importance 1,000 times. And as before, for (D6 - D8) we set $\lambda = 10\%$ and $t_0$ to be 35\% of the sample size and for (D7, D8) we use 500 subject pair resamplings to compute the matching threshold distance, $d_{t_\lambda}$. The results appear in Table~\ref{tab:convergence_guessing}.

\begin{table}[htp]
%\small
\centering
\begin{tabular}{lcc}
 & Guess Rate  & Average absolute arm\\ 
Design		& avg $\pm$ std. err. (\%) &  allocation difference (\%)\\ \hline
D1: Bernoulli 		& 50.14 $\pm$ 0.16		& 3.94					\\
D2: BCRD			& 55.53  $\pm$ 0.10		& 0				\\
D3: Biased 		& 62.32   $\pm$ 	0.12	& 0.71				\\
D4: Minimization	& 67.59  $\pm$ 0.12		& 0.28				\\

D4: KK14			&  53.97 $\pm$ 0.11		& 1.40				\\
D5: Naive 			& 54.11  $\pm$ 0.12		& 1.32				\\
D6: Stepwise 		& 54.32	 $\pm$ 0.12	& 1.19				\\
%D4*: KK14			& 56.07 $\pm$ 0.14	& \mdash					\\
%D5*: Naive 		& 56.38  $\pm$ 0.15		&	\mdash			\\
%D6*: Stepwise 		& 56.64$\pm$ 0.15 	&	\mdash				\\
\end{tabular}
\caption{Guess rate and arm imbalance results to two significant digits for 1,000 simulations of the convergence guessing strategy among the considered designs.}
\label{tab:convergence_guessing}
\end{table}

All of our sequential matching designs (D6-D8) perform about the same with a guess rate of 54\%, statistically significantly lower than the BCRD design (because in BCRD it is compulsory to have equal treatment arm allocations). It is interesting to note that Efron's Biased coin design which was motivated to be a simple means of minimizing investigator bias performs much worse than our sequential matching designs. However, our matching designs have marginally worse arm balance which reduces power and efficiency but this loss is more than compensated for by providing response-relevant matches and custom estimation as demonstrated in the previous section.

Even though the investigator can guess above chance in our design (and all other popular sequential designs), \qu{the idea that responsible investigators, even if they knew all the allocations to date, would spend their time playing games to try to guess a relatively complicated deterministic procedure... [is not] appealing} \citep{Begg1980}. This tampering is already nearly impossible in a simple multi-center setting \citep{Mcentegart2003}.

\subsection{No Time Trend in our Design}\label{sec:time_trend}

In any CARA design, there is a concern that allocations can trend with respect to the severity of the subject's response i.e. $t$ and $y$ exhibit a correlation \citep{Villar2018, Proschan2020}. Intuitively, this would not be the case in our CARA proposal as our design features completely random Bernoulli assignments for the first $n_0$ entrants and subsequently, either completely random Bernoulli assignments or matches which are the opposite allocation of subjects who were originally completely random Bernoulli assignments.

To test for a time trend, we run a simulation where we investigate the scenario of the vanilla dataset of Section~\ref{sec:simulations} at $n=100$ where the subjects are sorted in time by $f(x_{i1}, x_{i2}) + \mathcal{E}_i$, i.e. the response without the treatment effect. This sorting induces a strong monotonically increasing time trend (the subjects appear in order of severity). The simulation settings we choose are the most unfavorable to our CARA design. We employ the varied $\beta$'s (B2); correlated covariates (X2); our stepwise sequential weighted allocation algorithm (D8); the OLS estimator (E4) which is most prone to bias; and we employ the randomization test (T2) only because we know that the normal theory-based test has inflated size due to the asymptotics not engaging by $n=100$. We then ran 5,000 replicates of our design.

In the test of even probability of allocation to the treatment arm for each subject $t$, i.e. $H_0$: $p_t=50\%$ via a z-test, we found that no tests are rejected at $\alpha=5\%$ under Bonferroni correction. To confirm no time trend, we ran a logistic regression of the $w_i$'s on $t$ to test for a linear trend and a Cochran-Armitage test of the number of allocations to T and C at every subject versus the ordinal $t$ to test for a monotonic trend. Both tests were unable to reject a null hypothesis of no time trend (unshown). Further, a test of difference in the allocation probability generating process before and after $n_0$ failed to find a significant effect via a two-sample Kolmogorov-Smirnov test (unshown).

\subsection{Clinical Trial Data}\label{sec:clinical_data}

As in KK14, we examine the clinical trial data from a 12-week, multicenter, double-blind, placebo-controlled sequential clinical trial studying whether amitriptyline, an anti-depressant drug, can effectively treat painful bladder syndrome \citep{Foster2010}. The primary endpoint was whether their pain at least moderately improved (an incidence metric). Our methods are developed for regression endpoints so the response we consider is one of the investigators' secondary endpoints: change in pain after 12 weeks assessed on a 10-point Likert scale. The sample average amitriptyline effect estimate was -.4 on this scale with a 95\% confidence interval of [-1.00, .30] and a frequentist p-value of .29 (ibid, Table 2, row 1, page 1856) both constructed using the t-statistic (hence assuming the population model). This result either means the drug has no effect or the study was underpowered since the variation in the data was too large to detect its effect at its sample size of $n = 224$. 

To simulate our design, we first identify a subset of covariates from the original data that were measured for all the subjects: age, sex, race = White, race = Hispanic, level of education, level of employment, living with a partner, presence of sexually transmitted diseases, urinary tract infection, living with partner, as well as baseline measures of pain, urgency, frequency, problem index, symptom index, symptom inventory physical quality of life, mental quality of life, anxiety and depression for a total of 19 covariates. Peeking at the entire dataset, we note three observations (1) $R^2$ was about 24\% in an OLS regression without the treatment indicator, (2) the slope coefficients (after the covariates were normalized) illustrated only a few important variables and (3) there was a fair amount of collinearity between six of the baseline covariates (see their correlelogram, Figure~\ref{fig:correlelogram} in the Supporting Information). 

Observation (1) was noted by KK14 who conjectured that this $R^2$ would be sufficient for on-the-fly matching (without weighting the covariates) to perform better than the naive estimator $\YbarT - \YbarC$. Observation (2) is likely sufficient for our method presented in this paper to outperform KK14 and observation (3) indicates that the stepwise weighting algorithm will further improve our performance.

As in KK14, we simulate the subjects being dynamically allocated using our proposed CARA sequential design using many pseudo-datasets as replications. To do so, we first assume the subjects' order is not informative. Each sequential pseudo-dataset can then be a permutation of the order of the $n$ subjects' arrivals. In each of these replications, we employ our sequential design (D8) in the following way. The first $t_0$ subjects' assignments are retained. After $t_0$, we attempt to match according to our procedure. Half of the time, the matched subject in the actual clinical trial was assigned the opposite of the reservoir subject. But the other half of the time, the matched subject in the actual clinical trial was assigned the same as the reservoir subject. Within this replication, we retain the entering subject in the former and discard the entering subject in the latter (see KK14, Figure 2). In each replication, there is a different subset of the original data retained and we denote the size of this subset as $n_{rep}$.

In each pseudo-trial, we compute the estimate using the modified classic estimator of Equation~\ref{eq:our_estimator}. The performance of this estimator constitutes an apples-to-apples comparison to the analysis in \citet{Foster2010}. We do the comparison at the same sample size (using the same $n_{rep}$ subjects). We estimate the relative efficiency of our CARA design by computing the sample variance of our estimator divided by the sample variance of the naive estimator $\YbarT - \YbarC$. Since the focus is on the variance of the estimator (and not testing), the comparisons here are not affected by the observed anti-conservative sizing of the tests in Table~\ref{tab:size50}.

We further run replications of different subsample sizes that are independently and randomly drawn from the full sample size of 224 to explore our efficiency advantage over the naive estimator as sample size increases. The results averaged over 200 replications for each of subsamples of size $n = 50, 100, 150$ and all 224 are shown in Table~\ref{tab:clinical_trial_sim_results} where it is also compared to the results from KK14, Table 4, section (a). 

\begin{table}
\centering
\begin{tabular}{crccc}
																	&					& average			& average		& approx. sample \\
																	& purported $n$ 	& actual $n_{rep}$  		& efficiency 	& size reduction (\%) \\ \hline
\multirow[c]{4}{*}{(D6: KK14)} 		 								& 50				& 38.9				& 1.30			& 23.0 \\
  																	& 100				& 75.2				& 1.10			& 9.2 \\
  																	& 150				& 111.3			& 1.05			& 4.9 \\
  																	& 224 (all)			& 165.5			& 1.07			& 6.7 \\ \hline
\multirow[c]{4}{*}{(D8: stepwise)}  									& 50				& 39.3				& 1.23			& 18.8 \\
  																	& 100				& 75.3				& 1.15			& 12.7 \\
  																	& 150				& 111.7			& 1.13			& 11.6 \\
  																	& 224 (all)			& 164.3			& 1.10			& 9.2 \\ \hline
\end{tabular}
\caption{Clinical trial simulation results.}
\label{tab:clinical_trial_sim_results}
\end{table}

We see that for the average $n_{rep} > 40$, our design provides higher efficiency (and thus power) over the KK14 design that does not weight the covariates. The reason we do not see the advantage of our new procedure over KK14 at the small sample size is because our new design experiences noisy estimation of the 19 covariate weights. Poor estimates for weights perform worse than the equal weights of KK14. One may further ask why our stepwise CARA does not perform even better than shown given the impressive performance observed in Section~\ref{sec:simulations}. We believe this is due to a low upper bound on optimal performance as the $R^2$ of all covariates on response is only 24\%.

\begin{figure}[htp]
\centering
	\begin{subfigure}{7in}
		\hspace{-.1cm}\includegraphics[width=6.5in]{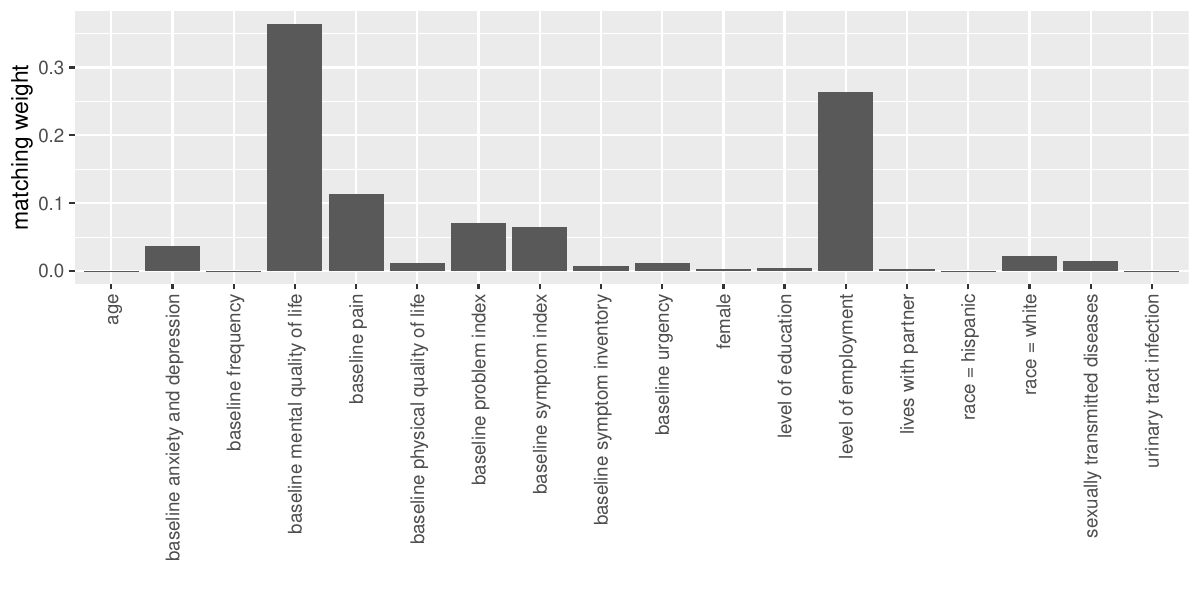}
		\caption{At $t = 18$}
		\label{fig:weightsa}
	\end{subfigure} \\
	\begin{subfigure}{5in}
		\hspace{-2cm}\includegraphics[width=6.5in]{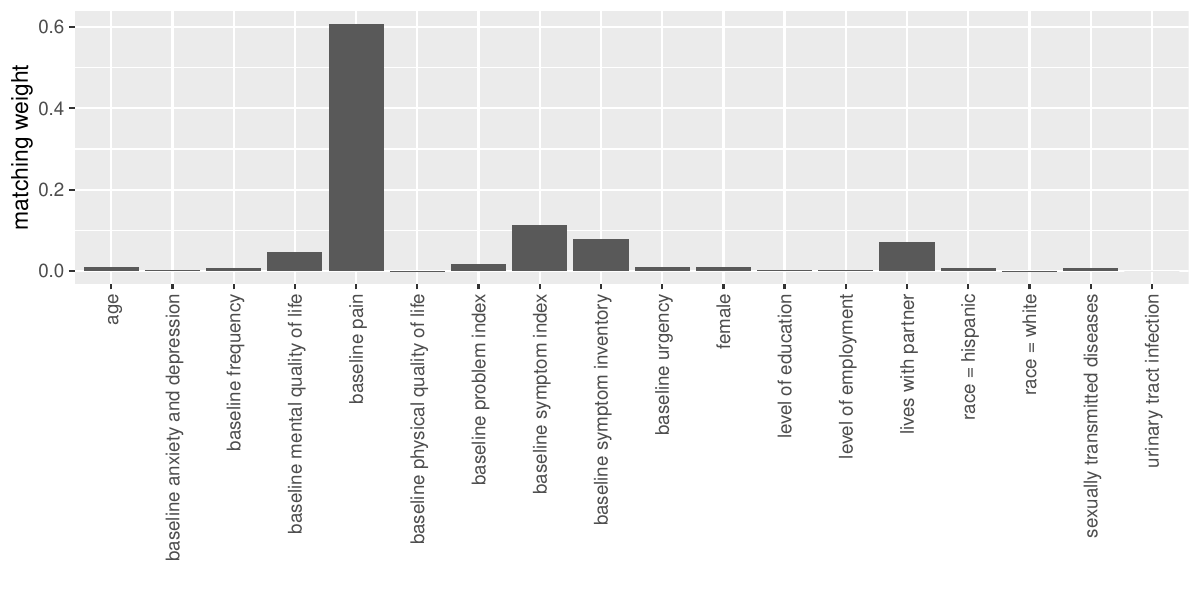}
		\caption{At $t = 168$}
		\label{fig:weightsb}
	\end{subfigure}
\caption{Normalized covariate weights estimated at (a) the beginning and (b) the end of a stepwise on-the-fly matching design for the clinical trial data.}
\label{fig:weights}
\end{figure}

Figure~\ref{fig:weightsa} shows our stepwise procedure's covariate weight estimates during a simulated trial with a subsample of 50 when it begins matching at $n = \ceiling{50 \times t_0 = 35\%} = 18$ and Figure~\ref{fig:weightsb} illustrates our stepwise procedure's covariate weight estimates at the end of a simulated trial with all subjects (in the run illustrated $n = 168$ due to the discarding).

By the end of the trial, the algorithm discovers that there is one covariate that dominates: baseline pain with a weight of about 60\% (Figure~\ref{fig:weightsb}). Once many subjects are observed, our design iteratively begins to match almost exclusively on this sole subject characteristic. This is the main reason the efficiencies are higher for our stepwise procedure when compared to KK14: matching on the very few covariates that truly are associated with the endpoint makes for better matches than matching on all 19 covariates indiscriminately. From the low-sample illustration of Figure~\ref{fig:weightsa}, we see that it erroneously chooses baseline mental quality of life and level of employment as the most important covariates; these are patient characteristics that are largely unassociated with the endpoint. As a result, the matches are poor quality. If we don't provide a sufficient $n$ to estimate the weights sufficiently accurately, our stepwise design cannot compete with even the indiscriminate covariate distance metric of KK14. This may suggest that in practice $t_0$ should be sufficiently large relative to the number of covariates $p$.

\section{Discussion}\label{sec:discussion}

We develop a sequential experimental design that makes use of both subjects' covariates and responses that are collected during the experiment. Our design matches similar subjects on-the-fly only if they are suitably similar. This iterative similarity assessment gauges the specific covariates that should be matched and to what degree by fitting collinearity-robust covariate models for the response. The ultimate result is treatment estimators with low variance and thus, higher power during testing. We offered two new designs of which we recommend for practice the stepwise on-the-fly covariate weighted matching design because it performs the best under the most general response model (i.e. with differential variable importance and collinearity among the variables) and does not perform appreciably worse under simpler models. We offered four testing variants and recommend in practice the exact test because it is appropriately sized (echoing Section~5 of KK14) with the OLS estimate employed as its internal test statistic as its power will be higher than employing the differences-in-means internal test statistic. We offer an \texttt{R} package \texttt{SeqExpMatch} which implements the designs explored in this paper most notably the new naive and stepwise designs. The software also provides estimation and testing for all four of the inferential settings herein (NC/PO, NC/NR, LC/PO and LC/NR).

The results in this article can be extended in various ways. The algorithm has tuning parameters which are chosen at default levels that worked well in our explorations. They include the number of individuals that are assigned to the reservoir before beginning to match, the distance metric that gauges distance between subjects (specifically how this metric's internal covariate weights are computed) and the maximum distance threshold that is deemed close enough for matching. For example, the first tuning parameter, the number of individuals that are assigned perforce to the reservoir initially, could depend on $n$ and $p$. The second tuning parameter, the dynamic cutoff for the distance function, could also depend on $n_R$ and $n - t$ in addition to the distances already observed. 

%At the end of the trial all subjects are assigned. A refinement one may consider is to take the final assignments and then use conventional methods to match the subjects post factum ignoring the matching structure that was created on-the-fly. Moreover, not performing a full bipartite matching will allow for the use of the convex combination estimators of Equations~\ref{eq:our_estimator} and \ref{eq:our_ols_estimator}. We tried this design strategy and found no improvement.

The setting of the paper can also be made more general by adapting the matching on-the-fly to experiments with $k > 2$ arms. We can also develop a consistent means of testing the null treatment effect hypothesis sequentially at $t < n$ as in \citet{Plamadeala2012}.

Some subset of the $p$ covariates may not be measured for the $t$th entering subject. We offer many options for this situation. First, it is possible to impute the missing covariate values using standard procedures and then run our algorithm as stated. We can also code missingness as dummy variables by covariate to be used as additional informative characteristics on which to match in case the missingness is not at random. Second, one can randomize this subject to the reservoir but bar this subject from serving as a match to a future subject. Third, our matching algorithm (as well as the algorithm in KK14) is flexible at each subject $t$ to measure distance and ultimately make a matching decision using any subset of available covariates. But if a subject is not matched to a reservoir subject at level $\lambda$ using the subset of covariates that are available, this subject is randomized but could be barred from future matches like explained previously.

Currently the weights are determined by normalizing the $R^2_j$'s from univariate linear regressions (or stepwise regressions to adjust for collinearity). This univariate relationship is not an assumption about the true response model, but rather a vehicle for computing sensible weights. We may be able to obtain more realistic weights by including treatment-covariate interactions and possibly splines to capture non-linearities. In unshown simulation, employing the $R^2_j$'s from univariate generalized additive models improves our results further (and this is an option in our software package).

Additionally, the response of the previously assigned subject may not be available when the entering subject must be assigned. We can alter our algorithm to update the weights in blocks such as in \cite{Zhou2018} or only reweight when the information becomes available.

We believe further studies can demonstrate gains in the performance of our designs through dynamic choice of the hyperparameter $\lambda$ e.g. as a function of the progress of the study $t / n$ and the number of current subjects in the reservoir $n_R(t)$. Through our extensive experience with simulations, our intuition is the gain herein is modest.

\backmatter

\section*{Acknowledgements}

We thank Katherine Propert for providing the clinical data used in Section~\ref{sec:clinical_data}. We would like to thank the associate editor and three referees for helpful suggestions that have improved the content and clarity of this work. This research was supported by Grant No. 2018112 from the United States-Israel Binational Science Foundation (BSF).

\section*{Supporting Information}

Web Appendix A containing additional figures and tables is available under the Paper Information link at the \textit{Biometrics} website, \texttt{http:www.tibs.org/biometrics}.

\bibliographystyle{biom}\bibliography{refs}

\begin{thebibliography}{}

\bibitem[\protect\citeauthoryear{Atkinson}{Atkinson}{1982}]{Atkinson1982}
Atkinson, A.~C. (1982).
\newblock Optimum biased coin designs for sequential clinical trials with
  prognostic factors.
\newblock {\em Biometrika} {\bf 69,} 61--67.

\bibitem[\protect\citeauthoryear{Atkinson and Biswas}{Atkinson and
  Biswas}{2005}]{Atkinson2005}
Atkinson, A.~C. and Biswas, A. (2005).
\newblock Adaptive biased-coin designs for skewing the allocation proportion in
  clinical trials with normal responses.
\newblock {\em Statistics in Medicine} {\bf 24,} 2477--2492.

\bibitem[\protect\citeauthoryear{Begg and Iglewicz}{Begg and
  Iglewicz}{1980}]{Begg1980}
Begg, C.~B. and Iglewicz, B. (1980).
\newblock A treatment allocation procedure for sequential clinical trials.
\newblock {\em Biometrics} pages 81--90.

\bibitem[\protect\citeauthoryear{Bertsimas, Johnson, and Kallus}{Bertsimas
  et~al.}{2015}]{Bertsimas2015}
Bertsimas, D., Johnson, M., and Kallus, N. (2015).
\newblock The power of optimization over randomization in designing experiments
  involving small samples.
\newblock {\em Operations Research} {\bf 63,} 868--876.

\bibitem[\protect\citeauthoryear{Bertsimas, Korolko, and Weinstein}{Bertsimas
  et~al.}{2019}]{Bertsimas2019}
Bertsimas, D., Korolko, N., and Weinstein, A.~M. (2019).
\newblock Covariate-adaptive optimization in online clinical trials.
\newblock {\em Operations Research} {\bf 67,} 1150--1161.

\bibitem[\protect\citeauthoryear{Bhat, Farias, Moallemi, and Sinha}{Bhat
  et~al.}{2020}]{Bhat2020}
Bhat, N., Farias, V.~F., Moallemi, C.~C., and Sinha, D. (2020).
\newblock Near-optimal ab testing.
\newblock {\em Management Science} .

\bibitem[\protect\citeauthoryear{Blackwell and Hodges}{Blackwell and
  Hodges}{1957}]{Blackwell1957}
Blackwell, D. and Hodges, Jr, J.~L. (1957).
\newblock Design for the control of selection bias.
\newblock {\em The Annals of Mathematical Statistics} pages 449--460.

\bibitem[\protect\citeauthoryear{Chandler and Kapelner}{Chandler and
  Kapelner}{2013}]{Chandler2013}
Chandler, D. and Kapelner, A. (2013).
\newblock Breaking monotony with meaning: Motivation in crowdsourcing markets.
\newblock {\em Journal of Economic Behavior \& Organization} {\bf 90,}
  123--133.

\bibitem[\protect\citeauthoryear{Chow and Chang}{Chow and
  Chang}{2008}]{Chow2008}
Chow, S. and Chang, M. (2008).
\newblock {Adaptive design methods in clinical trials - a review.}
\newblock {\em Orphanet journal of rare diseases} {\bf 3,} 11.

\bibitem[\protect\citeauthoryear{Cornfield}{Cornfield}{1959}]{Cornfield1959}
Cornfield, J. (1959).
\newblock Principles of research.
\newblock {\em American journal of mental deficiency} {\bf 64,} 240--252.

\bibitem[\protect\citeauthoryear{Efron}{Efron}{1971}]{Efron1971}
Efron, B. (1971).
\newblock {Forcing a sequential experiment to be balanced}.
\newblock {\em Biometrika} {\bf 58,} 403--417.

\bibitem[\protect\citeauthoryear{Fisher}{Fisher}{1925}]{Fisher1925}
Fisher, R.~A. (1925).
\newblock {\em Statistical methods for research workers}.
\newblock Edinburgh Oliver \& Boyd.

\bibitem[\protect\citeauthoryear{Foster, Hanno, Nickel, Payne, Mayer, Burks,
  Yang, Chai, Kreder, Peters, Lukacz, FitzGerald, Cen, Landis, Propert, Yang,
  Kusek, and Nyberg}{Foster et~al.}{2010}]{Foster2010}
Foster, H., Hanno, P., Nickel, J., Payne, C., Mayer, R., Burks, D., Yang, C.,
  Chai, T., Kreder, K., Peters, K., Lukacz, E., FitzGerald, M., Cen, L.,
  Landis, J., Propert, K., Yang, W., Kusek, J., and Nyberg, L. (2010).
\newblock {Effect of amitriptyline on symptoms in treatment na\"{\i}ve patients
  with interstitial cystitis/painful bladder syndrome.}
\newblock {\em The Journal of urology} {\bf 183,} 1853--1858.

\bibitem[\protect\citeauthoryear{Garthwaite}{Garthwaite}{1996}]{Garthwaite1996}
Garthwaite, P.~H. (1996).
\newblock Confidence intervals from randomization tests.
\newblock {\em Biometrics} {\bf 52,} 1387--1393.

\bibitem[\protect\citeauthoryear{Greevy, Lu, Silber, and Rosenbaum}{Greevy
  et~al.}{2004}]{Greevy2004}
Greevy, R., Lu, B., Silber, J.~H., and Rosenbaum, P. (2004).
\newblock Optimal multivariate matching before randomization.
\newblock {\em Biostatistics} {\bf 5,} 263--275.

\bibitem[\protect\citeauthoryear{Han, Enas, and McEntegart}{Han
  et~al.}{2009}]{Han2009}
Han, B., Enas, N., and McEntegart, D. (2009).
\newblock {Randomization by minimization for unbalanced treatment allocation}.
\newblock {\em Statistics in medicine} {\bf 28,} 3329--3346.

\bibitem[\protect\citeauthoryear{Horton, Rand, and Zeckhauser}{Horton
  et~al.}{2011}]{Horton2011}
Horton, J., Rand, D., and Zeckhauser, R. (2011).
\newblock {The online laboratory: conducting experiments in a real labor
  market}.
\newblock {\em Experimental Economics} {\bf 14,} 399--425.

\bibitem[\protect\citeauthoryear{Hu and Rosenberger}{Hu and
  Rosenberger}{2006}]{HuRosenberger2006}
Hu, F. and Rosenberger, W. (2006).
\newblock {\em The Theory of Response-Adaptive Randomization in Clinical
  Trials}.
\newblock John Wiley \& Sons, Inc.

\bibitem[\protect\citeauthoryear{Imbens and Rubin}{Imbens and
  Rubin}{2015}]{Imbens2015}
Imbens, G.~W. and Rubin, D.~B. (2015).
\newblock {\em Causal inference in statistics, social, and biomedical
  sciences}.
\newblock Cambridge University Press.

\bibitem[\protect\citeauthoryear{Kallus}{Kallus}{2018}]{Kallus2018}
Kallus, N. (2018).
\newblock Optimal a priori balance in the design of controlled experiments.
\newblock {\em Journal of the Royal Statistical Society: Series B (Statistical
  Methodology)} {\bf 80,} 85--112.

\bibitem[\protect\citeauthoryear{Kapelner and Krieger}{Kapelner and
  Krieger}{2014}]{Kapelner2014}
Kapelner, A. and Krieger, A. (2014).
\newblock Matching on-the-fly: Sequential allocation with higher power and
  efficiency.
\newblock {\em Biometrics} {\bf 70,} 378--388.

\bibitem[\protect\citeauthoryear{Kapelner, Krieger, Sklar, Shalit, and
  Azriel}{Kapelner et~al.}{2020}]{Kapelner2020}
Kapelner, A., Krieger, A.~M., Sklar, M., Shalit, U., and Azriel, D. (2020).
\newblock Harmonizing optimized designs with classic randomization in
  experiments.
\newblock {\em The American Statistician} pages 1--12.

\bibitem[\protect\citeauthoryear{Krieger, Azriel, Sklar, and Kapelner}{Krieger
  et~al.}{2020}]{Krieger2020}
Krieger, A.~M., Azriel, D., Sklar, M., and Kapelner, A. (2020).
\newblock Improving the power of the randomization test.
\newblock {\em arXiv preprint 2008.05980} .

\bibitem[\protect\citeauthoryear{McEntegart}{McEntegart}{2003}]{Mcentegart2003}
McEntegart, D.~J. (2003).
\newblock The pursuit of balance using stratified and dynamic randomization
  techniques: an overview.
\newblock {\em Drug Information Journal} {\bf 37,} 293--308.

\bibitem[\protect\citeauthoryear{Morgan and Rubin}{Morgan and
  Rubin}{2012}]{Morgan2012}
Morgan, K.~L. and Rubin, D.~B. (2012).
\newblock Rerandomization to improve covariate balance in experiments.
\newblock {\em The Annals of Statistics} pages 1263--1282.

\bibitem[\protect\citeauthoryear{Plamadeala and Rosenberger}{Plamadeala and
  Rosenberger}{2012}]{Plamadeala2012}
Plamadeala, V. and Rosenberger, W.~F. (2012).
\newblock Sequential monitoring with conditional randomization tests.
\newblock {\em The Annals of Statistics} {\bf 40,} 30--44.

\bibitem[\protect\citeauthoryear{Pocock and Simon}{Pocock and
  Simon}{1975}]{Pocock1975}
Pocock, S.~J. and Simon, R. (1975).
\newblock {Sequential treatment assignment with balancing for prognostic
  factors in the controlled clinical trial}.
\newblock {\em Biometrics} {\bf 31,} 103--115.

\bibitem[\protect\citeauthoryear{Proschan and Evans}{Proschan and
  Evans}{2020}]{Proschan2020}
Proschan, M. and Evans, S. (2020).
\newblock Resist the temptation of response-adaptive randomization.
\newblock {\em Clinical Infectious Diseases} {\bf 71,} 3002--3004.

\bibitem[\protect\citeauthoryear{Rosenberger and Lachin}{Rosenberger and
  Lachin}{2016}]{Rosenberger2016}
Rosenberger, W.~F. and Lachin, J.~M. (2016).
\newblock {\em Randomization in clinical trials: theory and practice}.
\newblock John Wiley \& Sons, second edition.

\bibitem[\protect\citeauthoryear{Senn}{Senn}{2013}]{Senn2013}
Senn, S. (2013).
\newblock Seven myths of randomisation in clinical trials.
\newblock {\em Statistics in Medicine} {\bf 32,} 1439--1450.

\bibitem[\protect\citeauthoryear{Student}{Student}{1931}]{Student1931}
Student (1931).
\newblock {The Lanarkshire milk experiment}.
\newblock {\em Biometrika} {\bf 23,} 398--406.

\bibitem[\protect\citeauthoryear{Sverdlov, Rosenberger, and Ryeznik}{Sverdlov
  et~al.}{2013}]{Sverdlov2013}
Sverdlov, O., Rosenberger, W.~F., and Ryeznik, Y. (2013).
\newblock Utility of covariate-adjusted response-adaptive randomization in
  survival trials.
\newblock {\em Statistics in Biopharmaceutical Research} {\bf 5,} 38--53.

\bibitem[\protect\citeauthoryear{Villar, Bowden, and Wason}{Villar
  et~al.}{2018}]{Villar2018}
Villar, S.~S., Bowden, J., and Wason, J. (2018).
\newblock Response-adaptive designs for binary responses: How to offer patient
  benefit while being robust to time trends?
\newblock {\em Pharmaceutical statistics} {\bf 17,} 182--197.

\bibitem[\protect\citeauthoryear{Wiens}{Wiens}{2005}]{Wiens2005}
Wiens, D.~P. (2005).
\newblock Robust allocation schemes for clinical trials with prognostic
  factors.
\newblock {\em Journal of statistical planning and inference} {\bf 127,}
  323--340.

\bibitem[\protect\citeauthoryear{Wu}{Wu}{1981}]{Wu1981}
Wu, C.-F. (1981).
\newblock On the robustness and efficiency of some randomized designs.
\newblock {\em The Annals of Statistics} {\bf 9,} 1168--1177.

\bibitem[\protect\citeauthoryear{Zhang, Hu, Cheung, Chan, et~al\mbox{.}}{Zhang
  et~al.}{2007}]{Zhang2007}
Zhang, L.-X., Hu, F., Cheung, S.~H., Chan, W.~S., et~al. (2007).
\newblock Asymptotic properties of covariate-adjusted response-adaptive
  designs.
\newblock {\em The Annals of Statistics} {\bf 35,} 1166--1182.

\bibitem[\protect\citeauthoryear{Zhou, Ernst, Morgan, Rubin, and Zhang}{Zhou
  et~al.}{2018}]{Zhou2018}
Zhou, Q., Ernst, P.~A., Morgan, K.~L., Rubin, D.~B., and Zhang, A. (2018).
\newblock Sequential rerandomization.
\newblock {\em Biometrika} {\bf 105,} 745--752.

\end{thebibliography}

\pagebreak
\appendix

\section{A}

\begin{figure}[htp]
\centering
\includegraphics[width=7.0in]{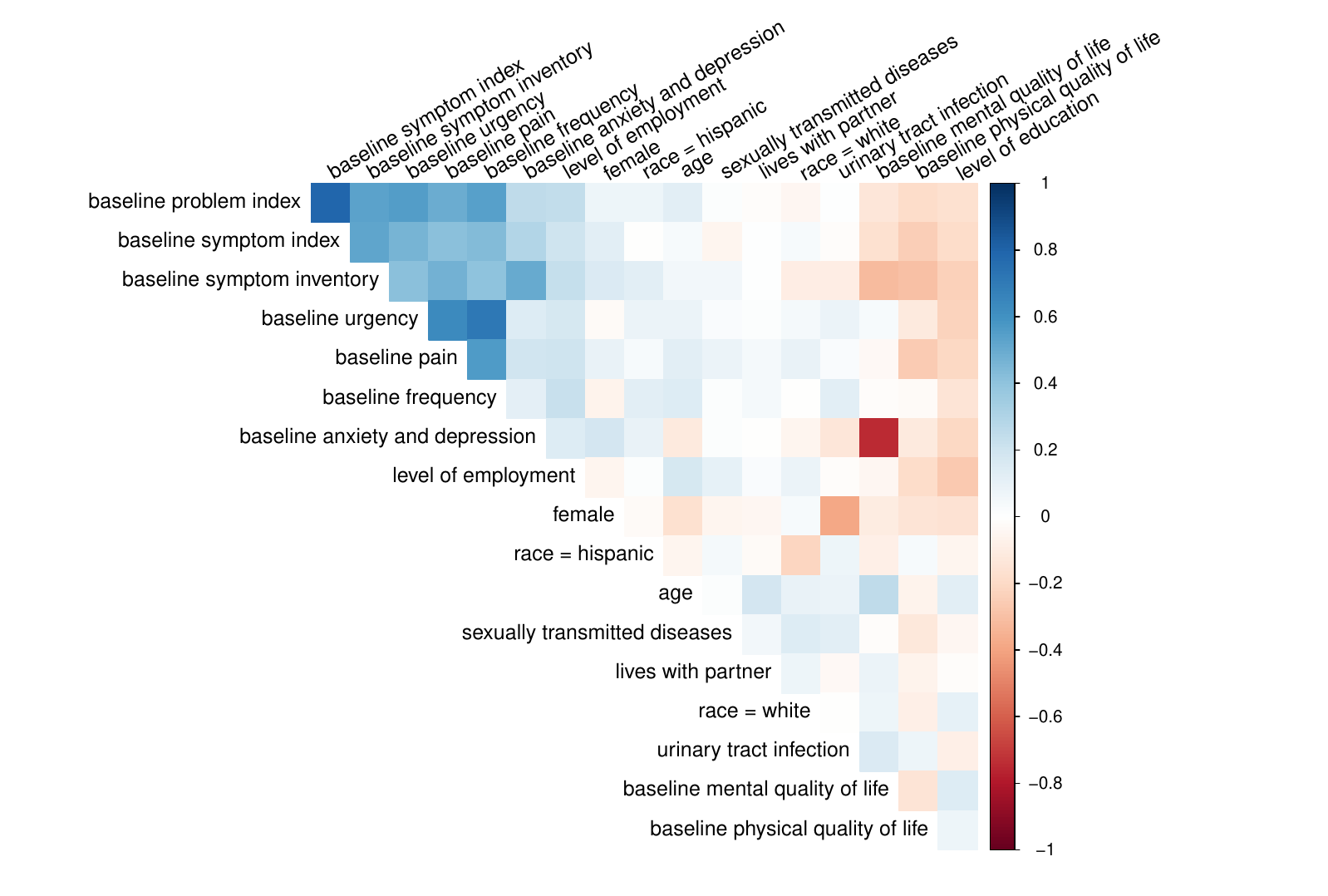}
\caption{Correlelogram for the 19 covariates considered in the clinical trial data.}
\label{fig:correlelogram}
\end{figure}

\begin{table}[htp]
%\small
\centering
\begin{tabular}{ccc|cc|cc}
											&															&				& \multicolumn{4}{c}{Betas} 																																	\\ 
											&															&				& \multicolumn{2}{c|}{B1: Uniform} & \multicolumn{2}{c}{B2: Varied}																\\ \cline{4-7}
											&															&				& \multicolumn{2}{c|}{Covariates } & \multicolumn{2}{c}{Covariates}\\
											&															&				& \multicolumn{2}{c|}{ Correlated?} & \multicolumn{2}{c}{Correlated?}\\
  											&   															&  			 & No				& Yes	 		& No				& Yes	 																												\\ \cline{4-7}
Design										&	Estimate												&	Test	 & \multicolumn{4}{c}{Power} 																														\\  \hline
 \multirow[c]{4}{*}{D1: Bernoulli}  		& \multirow[c]{2}{*}{E1: Classic} 						&	 NBT 		& .279 			& .222 		& .073 			& .071 \\
 											& 															& RAN  		& .292 			& .226 		& .078 			& .076 \\ \cline{2-7}
 											& \multirow[c]{2}{*}{E2: OLS}	 						&	 NBT 		& .814 			& .814 		& .401 			& .402 \\ 
 											& 															& RAN  		& .802 			& .814 		& .406 			& .413\\ \cline{1-7}
 \multirow[c]{4}{*}{D2: BCRD}  			& \multirow[c]{2}{*}{E1: Classic} 						&	 NBT 		& .283 			& .225 		& .073 			& .072 \\
 											& 															& RAN  		& .291 			& .228 		& .080 			& .075 \\ \cline{2-7}
 											& \multirow[c]{2}{*}{E2: OLS}	 						&	 NBT 		& .822 			& .821 		& .412 			& .404 \\ 
 											& 															& RAN  		& .826 			& .824 		& .416 			& .415\\ \cline{1-7}
 \multirow[c]{4}{*}{D3: Biased} 			& \multirow[c]{2}{*}{E1: Classic} 						&	 NBT 		& .278       	& .218 		& .078 			& .072 \\ 
 											& 															& RAN  		& .292 			& .225 		& .081 			& .076\\ \cline{2-7}
											& \multirow[c]{2}{*}{E2: OLS}	 						&	 NBT 		& .822       	& .812		& .408 			& .398 \\ 
 											& 															& RAN  		& .816      		& .819 		& .420 			& .413\\ \cline{1-7}
 \multirow[c]{4}{*}{D4: Minimization}     & \multirow[c]{2}{*}{E1: Classic} 						&	 NBT 		& .251  		& .164 		& .027			& .019 \\ 
 										     & 															& RAN  		& -- 			& -- 		& -- 			& --\\ \cline{2-7} 
 											& \multirow[c]{2}{*}{E2: OLS}	 						&	 NBT 		& .838  		& .847 		& .402 			& .399 \\ 
 											& 															 & RAN  		& -- 			& -- 		& --			& --\\ \cline{1-7} 
 \multirow[c]{4}{*}{D5: Atkinson} 		& \multirow[c]{2}{*}{E1: Classic} 						&	 NBT 		& .186       	& .096 		& .002 			& .002 \\ 
 											& 															& RAN  		& .604 			& .547 		& .158 			& .143\\ \cline{2-7}
											& \multirow[c]{2}{*}{E2: OLS}	 						&	 NBT 		& .823       	& .822		& .417 			& .415 \\ 
 											& 															& RAN  		& .827      		& .821 		& .418 			& .418\\ \cline{1-7}
 \multirow[c]{4}{*}{D6: KK14} 			& \multirow[c]{2}{*}{E3: Combined Classic}  			&	 NBT 		& .962 			& .943 		& .521 			& .479 \\ 
 											& 															& RAN  		& .957			& .937 		& .504 			& .468\\ \cline{2-7}  
 											& \multirow[c]{2}{*}{E4: Combined OLS}   				&	 NBT 		& .988 			& .988 		& .951 			& .953 \\ 
											& 															& RAN  		& .984 			& .983		& .944 			& .944\\ \cline{1-7} 
 \multirow[c]{4}{*}{D7: naive} 			& \multirow[c]{2}{*}{E3: Combined Classic}   			&	 NBT 		& .977 			& .980 		& .880 			& .708 \\ 
 											& 															& RAN  		& .977 			& .975 		& .868 			& .689\\ \cline{2-7}   
 											& \multirow[c]{2}{*}{E4: Combined OLS}  	 			&	 NBT 		& .994 			& .993 		& .991 			& .976 \\ 
 											& 															& RAN  		& .990 			& .988 		& .987 			& .970\\ \cline{1-7}  
 \multirow[c]{4}{*}{D8: stepwise} 		& \multirow[c]{2}{*}{E3: Combined Classic}   			&	 NBT 		& .982 			& .986 		& .900 			& .931 \\ 
	 										& 															& RAN  		& .979 			& .985 		& .894 			& .924\\ \cline{2-7}    
 											& \multirow[c]{2}{*}{E4: Combined OLS} 				&	 NBT		& .993 			& .994 		& .993 			& .992 \\ 
 											& 															& RAN  		& .990 			& .988 		& .989 			& .989\\ \hline
\end{tabular}
\caption{Experimental power (A1) when $n=100$ under the 120 varied simulation settings rounded to the nearest tenth of a percent.}
\label{tab:power100}
\end{table}

\begin{table}[htp]
%\small
\centering
\begin{tabular}{ccc|cc|cc}
											&															&				& \multicolumn{4}{c}{Betas} 																																	\\ 
											&															&				& \multicolumn{2}{c|}{B1: Uniform} & \multicolumn{2}{c}{B2: Varied}																\\ \cline{4-7}
											&															&				& \multicolumn{2}{c|}{Covariates } & \multicolumn{2}{c}{Covariates}\\
											&															&				& \multicolumn{2}{c|}{ Correlated?} & \multicolumn{2}{c}{Correlated?}\\
  											&   															&  			 & No				& Yes	 		& No				& Yes	 																												\\ \cline{4-7}
Design										&	Estimate												&	Test	 & \multicolumn{4}{c}{Power} 																														\\  \hline
 \multirow[c]{4}{*}{D1: Bernoulli}  		& \multirow[c]{2}{*}{E1: Classic} 						&	 NBT 		& .0503 			& .0496 		& .0464 			& .0490 \\
 											& 															& RAN  		& .0541 			& .0535 		& .0467 			& .0493 \\ \cline{2-7}
 											& \multirow[c]{2}{*}{E2: OLS}	 						&	 NBT 		& .0472 			& .0485 		& .0487 			& .0463 \\ 
 											& 															& RAN  		& .0522 			& .0525 		& .0523 			& .0526\\ \cline{1-7}
 \multirow[c]{4}{*}{D2: BCRD}  			& \multirow[c]{2}{*}{E1: Classic} 						&	 NBT 		& .0492 			& .0480 		& .0525 			& .0490 \\
 											& 															& RAN  		& .0538 			& .0511 		& .0505 			& .0486 \\ \cline{2-7}
 											& \multirow[c]{2}{*}{E2: OLS}	 						&	 NBT 		& .0497 			& .0481 		& .0505 			& .0486 \\ 
 											& 															& RAN  		& .0509 			& .0514 		& .0552 			& .0539\\ \cline{1-7}
 \multirow[c]{4}{*}{D3: Biased} 			& \multirow[c]{2}{*}{E1: Classic} 						&	 NBT 		& .0573       		& .0477 		& .0508 			& .0491 \\ 
 											& 															& RAN 			& .0591 			& .0484 		& .0499 			& .0512\\ \cline{2-7}
											& \multirow[c]{2}{*}{E2: OLS}	 						&	 NBT 		& .0481       		& .0455		& .0480 			& .0490 \\ 
 											& 															& RAN  		& .0500      		& .0496 		& .0515 			& .0561\\ \cline{1-7}
 \multirow[c]{4}{*}{D4: Minimization}     & \multirow[c]{2}{*}{E1: Classic} 						&	 NBT 		& .0179*  			& .0110* 		& .0105*			& .0050* \\ 
 										     & 															& RAN  		& -- 				& -- 			& -- 				& --\\ \cline{2-7} 
 											& \multirow[c]{2}{*}{E2: OLS}	 						&	 NBT 		& .0269*			& .0271* 		& .0240* 			& .0173* \\ 
 											& 															& RAN  		& -- 				& -- 			& --				& --\\ \cline{1-7} 
 \multirow[c]{4}{*}{D5: Atkinson} 		& \multirow[c]{2}{*}{E1: Classic} 						&	 NBT 		& .0016*       		& .0009* 		& .0001* 			& .0000* \\ 
 											& 															& RAN  		& .0544		 	& .0504 		& .0527 			& .0517\\ \cline{2-7}
											& \multirow[c]{2}{*}{E2: OLS}	 						&	 NBT 		& .0472       		& .0477		& .0509 			& .0515 \\ 
 											& 															& RAN  		& .0524      		& .0535 		& .0553 			& .0498\\ \cline{1-7}
 \multirow[c]{4}{*}{D6: KK14} 			& \multirow[c]{2}{*}{E3: Combined Classic}  			&	 NBT 		& .0611* 			& .0554 		& .0569 			& .0552 \\ 
 											& 															& RAN  		& .0537			& .0501 		& .0531 			& .0508\\ \cline{2-7}  
 											& \multirow[c]{2}{*}{E4: Combined OLS}   				&	 NBT 		& .0610* 			& .0615* 		& .0597* 			& .0571 \\ 
											& 															& RAN  		& .0537 			& .0539		& .0547 			& .0498\\ \cline{1-7} 
 \multirow[c]{4}{*}{D7: naive} 			& \multirow[c]{2}{*}{E3: Combined Classic}   			&	 NBT 		& .0585* 			& .0559 		& .0543 			& .0576 \\ 
 											& 															& RAN  		& .0517 			& .0481 		& .0537 			& .0532\\ \cline{2-7}   
 											& \multirow[c]{2}{*}{E4: Combined OLS}  	 			&	 NBT 		& .0619* 			& .0604* 		& .0548 			& .0559 \\ 
 											& 															& RAN  		& .0514 			& .0510 		& .0529 			& .0518\\ \cline{1-7}  
 \multirow[c]{4}{*}{D8: stepwise} 		& \multirow[c]{2}{*}{E3: Combined Classic}   			&	 NBT 		& .0581* 			& .0624* 		& .0574 			& .0546 \\ 
	 										& 															& RAN  		& .0510 			& .0507 		& .0495 			& .0567\\ \cline{2-7}    
 											& \multirow[c]{2}{*}{E4: Combined OLS} 				&	 NBT		& .0627* 			& .0557 		& .0570 			& .0566 \\ 
 											& 															& RAN  		& .0523 			& .0510 		& .0555 			& .0500\\ \hline
\end{tabular}
\caption{Experimental size (A2) when $n=100$ under the 120 varied simulation settings rounded to the nearest hundredth of a percent. Asterisks indicate the test is likely not properly sized as the null hypothesis that size equals the purported 5\% tested at $\alpha = 5\%$ with Bonferroni corrections for multiple hypothesis testing was rejected (the rejection is less than 0.04231 and greater than 0.05769).}
%> qnorm(0.05/2/120)
%[1] -3.529296
%> .05 + 3.529296 * sqrt(0.05*.95 / 10000)
%[1] 0.05769192
%> .05 - 3.529296 * sqrt(0.05*.95 / 10000)
%[1] 0.04230808
\label{tab:size100}
\end{table}

\end{document}